\documentclass[rmp,aps,twocolumn,amsmath,amssymb,superscriptaddress,longbibliography,10pt]{revtex4-1}

\usepackage{graphicx}
\usepackage{dcolumn}
\usepackage{bm}
\usepackage{psfrag}
\usepackage{mathrsfs} 
\usepackage{multirow} 

\DeclareFontFamily{OMS}{oasy}{\skewchar\font48 }
\DeclareFontShape{OMS}{oasy}{m}{n}{%
         <-5.5> oasy5     <5.5-6.5> oasy6
      <6.5-7.5> oasy7     <7.5-8.5> oasy8
      <8.5-9.5> oasy9     <9.5->  oasy10
      }{}
\DeclareFontShape{OMS}{oasy}{b}{n}{%
       <-6> oabsy5
      <6-8> oabsy7
      <8->  oabsy10
      }{}
\DeclareSymbolFont{oasy}{OMS}{oasy}{m}{n}
\SetSymbolFont{oasy}{bold}{OMS}{oasy}{b}{n}

\DeclareMathSymbol{\smallleftarrow}     {\mathrel}{oasy}{"20}
\DeclareMathSymbol{\smallrightarrow}    {\mathrel}{oasy}{"21}
\DeclareMathSymbol{\smallleftrightarrow}{\mathrel}{oasy}{"24}

\newcommand{\mytensor}[1]{\overset{\scriptscriptstyle\smallleftrightarrow}{#1}}

\newcommand{\bra}[1]{\langle #1|}

\newcommand{\MyFourTerm}{x\gg \Delta x \gg \frac{\lambda}{4\pi} \gg \frac{\Delta \lambda}{4 \pi}}

\begin{document}

\title{An Ahistorical Approach to Elementary Physics}

\author{B.C. Regan}
\email{regan@physics.ucla.edu}
\affiliation{Department of Physics and Astronomy, University of California, Los Angeles, California 90095, U.S.A.}
\affiliation{California NanoSystems Institute, University of California, Los Angeles, California 90095, U.S.A.}

\date{\today}

\begin{abstract}
A goal of physics is to understand the greatest possible breadth of natural phenomena in terms of the most economical set of basic concepts. However, as the understanding of physics has developed historically, its pedagogy and language have not kept pace.  This gap handicaps the student and the practitioner, making it harder to learn and apply ideas that are `well understood', and doubtless making it more difficult to see past those ideas to new discoveries.   Energy, momentum, and action are archaic concepts representing an unnecessary level of abstraction. Viewed from a modern perspective, these quantities correspond to wave parameters, namely temporal frequency, spatial frequency, and phase, respectively. The  main results of classical mechanics can be concisely reproduced by considering waves in the spacetime defined by special relativity. This approach unifies kinematics and dynamics, and introduces inertial mass not via a definition, but rather as the real-space effect of a reciprocal-space invariant. 

\end{abstract}


\maketitle


\section{Introduction}
Physics is customarily taught following the same chronological order in which it was developed.  Students first learn the classical mechanics of Galileo and Newton, then the classical electricity and magnetism of Faraday and Maxwell, and then more modern topics such as special relativity and quantum mechanics.  This approach has pedagogical advantages, for it features a steady progression of decreasing intuitiveness and increasing mathematical sophistication.  

However, the opportunity cost is high. The chronological approach  misses the chance to present the subject from the clearest, most logically consistent perspective, weighed down as it is with so much historical baggage.  Repeatedly the student is asked to unlearn concepts previously presented as foundational in favor of new ones, which in turn are overthrown in the next course.

Not only is physics not taught in a sequence that emphasizes our best, present understanding,  this perspective is never presented. The standard physics curriculum never gets around to saying, ``if we were starting from scratch, this is how we would do it.'' While classical physics is known to be only an approximation of more modern theories, we persist in using its dated vocabulary, even in, for example, quantum mechanical situations,  where it is ill-suited to the purpose.  Of course, we experience the world as classical beings, and thus benefit from an understanding of the approximate theories.  But it is possible, and indeed easier, to access and admire the full power of the approximate theories if they are not cloaked in classical language, which was developed well before the underlying and unifying ideas were discovered.

The goal here is to present a formulation of classical mechanics that is concise and more closely based on our current, best understanding of physics as a whole. Unlike the chronological approach, this perspective does not require extensive revision as more advanced subjects are introduced. With only two main ideas, both already part of the undergraduate physics curriculum, we can derive most of classical mechanics (the exception being gravity, which even according to our best understanding requires special treatment), and lay a firm foundation upon which more advanced topics (e.g. electricity and magnetism, quantum mechanics) can be based. 

No new predictions result.  We cannot calculate anything that was previously incalculable.  In fact, much of our presentation reiterates what has already been written elsewhere.  For instance, it has been noted previously that, had quantum mechanics been discovered first, surely Planck's constant would have been chosen to be unity \cite{2017Sakurai,2004Abers}, and that quantum mechanics subsumes classical mechanics \cite{2005Ogborn}. However,  the shape and consequences of an ahistorical development of classical mechanics seem to be relatively unexplored. As we will show, the logical underpinnings of classical mechanics can be recast by exploiting the developments of the early twentieth century.  This route abbreviates the lexicon that other physics theories subsequently adopt --- by not starting with Newton, we discover that much of his vocabulary is unnecessary. It also provides a satisfying simplification, one that removes unnecessary definitions, convoluted developments, and unmotivated postulates. The resulting edifice offers some insight, and the pleasure that comes from seeing old ideas from a new perspective \cite{1948Feynman}.

\section{Spacetime}\label{sec:spacetime}

The spacetime of special relativity serves as the arena. We follow the customary development of the subject \cite{1992Taylor,1999Jackson,2002Goldstein}, postulating that the rules of physics are the same when viewed from any inertial reference frame, and that, in particular, all inertial observers observe the same value $c$ for the speed of light in vacuum. Applying this idea to simple physical systems (e.g. a pulsed light source with a mirror on a train) gives, via basic geometry, the Lorentz contraction and time dilation effects.  From these we derive the Lorentz transformations and the idea of an invariant, the spacetime interval $s$, which is seen to have the same value by all inertial observers.  In a standard 4-vector notation with $x^\mu  = (ct, \mathbf{r})$ we have,
\begin{equation}\label{eq:ssquared}
x_\mu x^\mu=c^2 t^2 - \mathbf{r}^2=s^2,
\end{equation}
where $x^\mu$ indicates  a location, or `event', in spacetime in terms of its time $t$ and its position $\mathbf{r}$.  We have chosen to put the $c$ with the time $t$, so that every term in this expression has the dimension of length-squared. 

Defining the speed of light $c$ allows us to describe events using just one dimensioned unit, and to more easily treat spacetime as a unified whole.  Physicists working with very large length scales, e.g. astronomers, tend to measure distances using a time unit: the light-year.  Others are generally better off measuring times in distance units, simply because short distances have a greater number of familiar landmarks.  Outside of the ultrafast community, an angstrom and a fermi are more recognizable as atomic and nuclear size scales, respectively, than their near counterparts the attosecond and the yoctosecond.

To lay the groundwork for future developments, it is important to emphasize here that, given the strong geometric basis for Eq.~\ref{eq:ssquared}, it would never occur to anyone to measure its right-hand side in terms of a new, dimensionful unit. That the customary units designating time $(t)$ and place $(\mathbf{r})$ differ by a factor of $c$ is already regrettable.  Introducing yet another unit specifically to describe the interval $s$  would only further muddle what ought to be a straightforward geometric (i.e. pseudo-Pythagorean) relationship. But, as we will see, just such a muddling occurs when the dynamic counterpart of Eq.~\ref{eq:ssquared} is presented in its traditional form. 


\section{Spacetime contains waves}\label{sec:waves}

We postulate that spacetime contains objects that we will, in a convenient shorthand, refer to as waves. (For further discussion of this choice, see Appendix~\ref{particles}.) In advanced treatments these objects might be described more carefully as, for example, fields or wave functions, but for our present purposes such specificity is unnecessary.  In fact, we do not even need to specify a particular wave equation, only that these waves have a phase $\phi$ of the form
\begin{equation} \label{eq:phi}
-\phi = f t - \mathbf{k} \cdot \mathbf{r},
\end{equation}
as is found in the solutions $u(\phi)$ to the generic wave equation 
\stepcounter{equation}
\begin{equation}\label{eq:genericwave}
-\frac{1}{v^2}\frac{\partial^2 u}{\partial t^2} + \nabla^2 u =0,\tag{\theequation a}
\newcounter{wavecounter}
\setcounter{wavecounter}{\theequation}
\end{equation}
with the phase velocity $v\equiv f/|\mathbf{k}|$.  For our present purpose of reformulating classical mechanics, the solutions of interest are wave packets that are well-localized in real and reciprocal space, e.g. Gaussian functions viewed at a zoom level such that their widths are negligible. This second assumption (waves) clearly has deep connections to the first (special relativity): Eqs.~\ref{eq:genericwave} and \ref{eq:ssquared} are structurally similar,  Eq.~\ref{eq:genericwave} is Lorentz invariant, and Eq.~\ref{eq:genericwave} does not  permit instantaneous action-at-a-distance (unlike, e.g. diffusion-type equations).  

The phase  $\phi(x^\mu)$ is a number (defined here with an overall minus sign for consistency with a historical convention to be introduced later) associated with a particular event $x^\mu = (ct,\mathbf{r})$ in spacetime. It can be determined by a counting operation, and thus all inertial observers agree upon its value \cite{1999Jackson,1933Dirac,1971Wichmann,1971Symon}.  Therefore $\phi$ is a Lorentz invariant, and it can be written $\phi(x^\mu) = - k_\mu x^\mu$, where we have defined a new 4-vector $k^\mu \equiv (f/c, \mathbf{k})$.  

Since both $s^2=x_\mu x^\mu$ (Eq.~\ref{eq:ssquared}) and $\phi= -k_\mu x^\mu$  (Eq.~\ref{eq:phi}) are Lorentz invariant,  $x^\mu$ and $k^\mu$ must have the same behavior under a Lorentz transformation. These transformation properties then ensure that the 4-vector product $k_\mu k^\mu$ is also a Lorentz invariant. Thus
\begin{equation}\label{eq:goodeinstein}
k_\mu k^\mu=(f/c)^2-\mathbf{k}^2=k_C^2,
\end{equation}
where the quantity $k_C$, which at this point is just a number seen to have the same value by all inertial observers, will shortly be identified as the Compton wavenumber.  

Equation~\ref{eq:goodeinstein} is the reciprocal-space, or Fourier-space, analog of Eq.~\ref{eq:ssquared}.  Although we have not yet introduced energy, momentum, or mass, Eq.~\ref{eq:goodeinstein} contains the physical content of Einstein's famous energy-momentum relation. Multiplying through by $(h c)^2$, where $h$ is Planck's constant,  on both sides of Eq.~\ref{eq:goodeinstein} puts this equation in its traditional form:
\begin{equation}\label{eq:badeinstein}
 p_\mu p^\mu \,c^2 = E^2-p^2 c^2 = m^2 c^4,
\end{equation} 
where we have defined the four vector $p^\mu=(E/c,\mathbf{p})$, as well as three `new' physical quantities: the energy $E \equiv hf$, the momentum $\mathbf{p} \equiv h \mathbf{k}$, and the invariant (or rest) mass $m\equiv h k_C/c$.

\begin{table*}
\begin{tabular}{cc|c|cc|c|cr@{$\,=\,$}l}
\multicolumn{2}{c|}{kinematic}& &\multicolumn{6}{c}{dynamic}\\
\hline
\multicolumn{2}{c|}{\multirow{2}{*}{Galileo}}&Lagrange&\multicolumn{6}{c}{\multirow{2}{*}{Newton}}\\
\multicolumn{2}{c|}{}&Hamilton&\multicolumn{6}{c} {}\\
\hline

\multicolumn{2}{c|}{real space} & Fourier & \multicolumn{2}{c|}{reciprocal space} & $\times h$& \multicolumn{3}{c}{traditional}\\
\hline\hline
time&$ct$&$\Longleftrightarrow$&temporal frequency&$f/c$&Planck-Einstein&energy&$E/c$&$hf/c$\\
position&$\mathbf{r}$&$\Longleftrightarrow$&spatial frequency&$\mathbf{k}$&de Broglie&momentum&$\mathbf{p}$&$h\mathbf{k}$\\
\hline
\multicolumn{8}{c} {Pythagoras, Einstein, Minkowski} \\
\hline
interval&$s$&$\Longleftrightarrow$&Compton wavenumber&$k_C$&Compton&mass&$m c$&$h k_C$\\
\hline\hline
event & $x^\mu$ & & wave&$k^\mu$ & & object&\multicolumn{2}{c}{$p^\mu$}\\
\end{tabular}

\caption{Every event in real space has three distinct properties: a time, a position, and an interval. The constant `$c$' accommodates the pseudo-Pythagorean relationship between these properties by putting them in the same units.  The constant `$h$', on the other hand, conceals the Fourier relationship between the kinematic and the dynamic quantities by taking them out of reciprocal units.   As for the dynamic quantities, every object has an energy, a momentum, and a mass, and every wave has a frequency, a spatial frequency (or wavevector), and a Compton wavenumber\footnote{Time, position, frequency, and spatial frequency can all have well-defined values (i.e. with negligible spread) simultaneously in the classical limit. See Appendix~\ref{particles}.}. The `object'  and the `wave' pictures give physically equivalent descriptions of the dynamic variables.  Because `mass' is just another name for the Compton wavenumber, which is a derived property of waves in spacetime, we can say that objects have mass because they are waves.}
\label{table:dictionary}
\end{table*}

These three relationships (Table~\ref{table:dictionary}) between wave parameters and their corresponding classical variables were discovered and introduced separately through the Planck-Einstein relation (1900--1905), the de Broglie hypothesis (1924), and the aforementioned Compton formula (1923), respectively. Typical pedagogy does not apply the Planck-Einstein relation to massive objects, and connects these three ideas only to the extent that they were all instrumental to the development of quantum mechanics.  In particular, it does not present the unifying, geometric picture implied by Eq.~\ref{eq:goodeinstein}.

The path to Eq.~\ref{eq:goodeinstein} as presented here is fast and direct: find the invariant length in real space,  introduce the Fourier-conjugate variables, and find the equivalent invariant length in reciprocal space.  In comparison, the traditional route to Eq.~\ref{eq:badeinstein}  is lengthy and circuitous. One first elevates momentum conservation to the status of a foundational principle.  Then, working backwards from Newton's laws and employing the work-energy theorem, one reverse-engineers definitions of momentum and energy that satisfy this principle \cite{1992Taylor,2004Thornton,2014Kleppner,1985Eisberg}. With a little more effort, one can obtain the same result with momentum and energy conservation alone \cite{1999Jackson,1971Symon}. These arguments are subtle (it is not immediately obvious that momentum is something other than merely that-quantity-which-is-conserved) and nearly circular (they justify the exact law by appeal to the inexact).  These traditional approaches also rely on an unnecessarily large number of axioms. As we will see in Section~\ref{sec:stationary},  we can derive energy conservation, momentum conservation, and Newton's laws all from Eq.~\ref{eq:phi}. 

Furthermore, getting to the traditional Einstein energy-momentum relation (Eq.~\ref{eq:badeinstein}) via reciprocal space reveals that the final step --- multiplying through by $(h c)^2$ --- has no physical content, and in fact buries some valuable ideas.  First, the Fourier-transform relationship between the conjugate pairs $(t,f)$ and $(\mathbf{r},\mathbf{k})$ is obscured in the `new' units.  Second, mass has been defined such that it agrees dimensionally with neither $E$ nor $p$, thus hiding what would otherwise be a straightforward pseudo-Pythagorean relationship in reciprocal space. In the language of Taylor and Wheeler's parable of the surveyors \cite{1992Taylor}, we are measuring north-south distances in joules, east-west distances in kg$\cdot$m/s, and the distance between two points in kilograms. Thus Eq.~\ref{eq:badeinstein}, one of the most famous equations in all of physics, has not just three units (in the sense of meters and feet), but three different \emph{dimensions} (in the sense of joules and kilograms) for its three terms, and could not more thoroughly obfuscate its geometric content.  It is small wonder that physics is considered an arcane discipline. 

Equation~\ref{eq:badeinstein} might seem to have more content than Eq.~\ref{eq:goodeinstein}, or at least to be more intuitive. In fact, energy and momentum are intuitive only to those who have been suitably trained --- one might say indoctrinated --- in their use.  These quantities are actually relatively abstract, since they cannot be measured  directly, even in principle. The lack of experimental access is a substantial failing. To appreciate the advantages of explicitly considering the means of measurement, recall that a key advance in the development of special relativity was these deliberate definitions: time is what clocks measure, and distance is what rulers measure.  What then, are the corresponding operational definitions \cite{2017Hecht} of energy and momentum?  Such definitions do not exist; we do not have meters for either energy or momentum.  It is an interesting exercise to ponder  how these quantities are determined experimentally in various  situations. For mechanics problems, we have only three basic measurement operations: counting, measuring position (with rulers), and measuring time (with clocks). Some combination of these three basic  measurements determine other physical quantities : frequency (both spatial and temporal), angle, velocity, acceleration,  energy, momentum, torque etc.  For example, to determine the momentum of a football, we would determine its mass and velocity separately, and then do a simple calculation.  We might weigh the football with a scale, thereby determining its mass via position (and an auxiliary determination of the acceleration due to gravity), i.e. the deflection of the spring. And we might use clocks  separated by a pre-measured distance to determine the velocity. 

In the list of `physical quantities' just given, the first two items, spatial and temporal frequency, are in a different category than the others.  They are not less fundamental than distance and time, in that distance and time cannot be defined without reference to frequency.  To be useful, clocks and rulers must contain some repeating unit.  A `clock' must tick more than once, and `rulers' must either be subdivided into sub-units, or be applicable in multiple instances (e.g. laid end-to-end).  Thus we cannot measure time without access to a frequency, nor can we measure distance without access to a spatial frequency. While the circularity of these base definitions might puzzle logicians and philosophers, this lexical issue causes no difficulty for physicists in practice.  It also emphasizes the symmetry (in the sense of a lack of precedence) between the real-space (i.e. spacetime-domain) and the reciprocal-space (i.e. frequency-domain) variables.\footnote{Since number is determined by counting, from a physics standpoint we consider it to be well-defined. Mathematically its definition is just as circular --- counting determines number, and number is what counting determines --- but we view mathematics as being on a firmer logical footing than physics. We understand number better than we understand time. 

Thus number is preeminent among the measurable quantities, as can be seen from the current (circular) statement defining the units of time and frequency in terms of number. With the ground-state hyperfine splitting of the cesium-133 atom as the standard oscillator, by definition one second is the time required to complete a number (9,192,631,770) of cycles, and the oscillator's frequency in hertz is that same number \cite{2016Stenger}. 

While classical mechanics treats both time and distance as measurable, more advanced physics makes the preeminence of number manifest. Quantum mechanics treats time as a parameter while maintaining position as an observable. Quantum field theory treats both time and position as parameters, which leaves number as the only measurable quantity remaining.}

Compared to energy and momentum, frequency and spatial frequency are much less abstract, in that they can be measured directly.  A frequency is measured by counting cycles in a known time.  A spatial frequency is measured by counting cycles in a known distance.  Of course, practical considerations make such direct measurements hopeless for macroscopic objects such as footballs (see note at the end of Section~\ref{sec:HistoricalPerspective}).

Replacing the archaic concepts of energy and momentum (Eq.~\ref{eq:badeinstein}) with frequency and spatial frequency (Eq.~\ref{eq:goodeinstein}), respectively, anticipates questions that can arise at an elementary level and yet are difficult to answer in the old language. The modern development avoids the question, ``how can photons have energy and momentum, if they do not have mass?''   Since a photon has zero mass (kg, in SI units) and non-infinite velocity (m/s), it is legitimate to ask how it can have non-zero energy (kg$\cdot$m$^2/$s$^2$) and momentum (kg$\cdot$m$/$s).  The answer, we see here, is that energy and momentum are better thought of as representing (inverse) time and length scales (i.e. frequencies), respectively, which the photon does have.  Mass too corresponds to a length scale, and for the photon this length happens to be infinite.  Mass was given its own, dimensioned unit long before the connection between mass and length was understood, and applying the historical language to a modern concept such as the photon leads to this inconsistency.

Thus, assuming waves in the spacetime of special relativity gives us mass. In other words, objects have mass \emph{because} they are waves (Table~\ref{table:dictionary}).  This statement is exactly counter the orthodox presentation --- usually the wave nature of a massive object is considered to arise \emph{despite} its mass. The conflict in the concept of `wave/particle duality' that is so often emphasized is partly an illusion.  

Having arrived at the concept of mass via this route (i.e. Fourier conjugate variables) further encourages us to re-examine the various classical conceptions of mass.  Historically mass has been considered ``a quantity of matter'', a disinclination to accelerate (i.e. inertia), or a measure of participation in the gravitational interaction \cite{2010Hecht}.  These ideas date back to Newton and have been continuously refined in the intervening centuries, but none of them is entirely satisfactory  \cite{2005Hecht,2010Hecht,2012Coelho}.  As we will show shortly, inertial mass does not need to be introduced as a postulate, since the inertial effect can be derived from other (wave) axioms.  Moreover, instead of having inertia defined in relation to the positions of the ``fixed stars'', as Mach would have it, inertia is defined in relation to positions in reciprocal space, a place much less remote. The inertia of a mass is not ``an effect due to the presence of all other masses'' \cite{2005Pais}, but rather a real-space effect arising from an invariance constraint (namely Eq.~\ref{eq:goodeinstein}) in reciprocal space. Distant, massive objects are not required;  Eq.~\ref{eq:goodeinstein} and the equations that follow (Section~\ref{sec:stationary}) suffice to derive the inertial effect. 

Thus we can take Eq.~\ref{eq:goodeinstein} to \emph{define} mass: mass is the invariant reciprocal length associated with a wave of frequency $f$ and wavevector $\mathbf{k}$ (times $h/c$ in traditional units).  This definition is exactly analogous to that of the interval: the interval is the invariant length associated with an event at time $t$ and position $\mathbf{r}$. Both definitions are closely tied to the structure of spacetime, and neither relies on the concept of `force', which is both difficult to define and almost ignored by advanced physics theories (e.g. quantum mechanics and quantum field theory).

The modern, wave-based approach further suggests a possible strategy for addressing a long-standing open problem, namely implementing an operational definition of mass \cite{2005Hecht}.  Intervals are generally determined with rulers and clocks (Eq.~\ref{eq:ssquared}), and, in the specific case of a purely time-like separation, an interval can be measured with just a clock.  A mass is the reciprocal-space analog of a purely time-like separation (Eq.~\ref{eq:goodeinstein}), which suggests that a mass might be determined by measuring its Compton frequency.  In fact, the use of `rocks as clocks' has been proposed \cite{2013Lan}, although the idea is not yet well understood \cite{2013Pease,2012Wolf,2013Schleich}.

\section{Stationary phase}\label{sec:stationary}

It is hardly an exaggeration to say that the rest of physics, excepting thermodynamics, will be built around the phase $\phi$.  One more postulate --- that $\phi$ is invariant under `gauge' (or phase) transformations --- will be used to introduce interactions into the theory.  Invariance under local $U(1)$ gauge transformations gives the electromagnetic interaction.  Invariance under local $SU(2)$ and $SU(3)$ gauge transformations leads to the weak and strong interactions respectively \cite{2010Griffiths,1991Cheng,1984Halzen}.  

Postponing these more advanced topics, we can derive the results of classical mechanics by invoking an extremum principle: the principle of stationary phase. We postulate that  classical trajectories or paths in spacetime are determined by the condition that $\phi$ is stationary along the trajectory \cite{1967Dirac,1977Cohen-Tannoudji,1995Peskin,2014Commins,2017Sakurai}, i.e. that $\delta \phi = 0$.  Trajectories not satisfying this condition are suppressed by destructive interference with neighboring trajectories; minute changes in the trajectory give an amplitude of the opposite sign. This idea, which is motivated by the wave picture, ultimately leads to the path integral formulation of quantum mechanics \cite{1948Feynman,1964Feynman,2010Feynman,2002Styer,1980Itzykson}.  Classically it is known as the principle of stationary action \cite{2006Ogborn} (or, in some nomenclatures, Hamilton's principle \cite{1964Feynman,1976Landau}), and it is generally presented as a postulate that has no underlying physical motivation, other than that it gives the desired equations of motion \cite{2004Thornton,2002Goldstein,1976Landau,1967Sakurai,1995Peskin,1967Dirac,1964Feynman,2010Feynman,1980Itzykson,1968Schiff}. Unlike the Planck-Einstein relation, the de Broglie hypothesis, and the Compton formula (Table~\ref{table:dictionary}), the relation $S=h \phi$ that connects the wave parameter ($\phi$, the phase) and the classical quantity ($S$, the action) is not well known, with some authors declining to state it clearly when the opportunity arises \cite{1995Peskin,2017Sakurai}. 

Having established that considerations of interference motivate the principle, we can derive the classical equations of motion by considering the phase $\phi$ as it develops along a path, and then applying the calculus of variations.  Then the derivation, based on the `modified Hamilton's principle', is standard \cite{2004Thornton,2002Goldstein,1976Landau}, but for the fact that the phase $\phi$, the frequency $f$, and wavevector $\mathbf{k}$ are taking the places of the action $S=h\phi$, the energy $E=h f$ (or, better, the Hamiltonian $H= h f$), and the momentum $\mathbf{p}=h\mathbf{k}$, respectively (Table~\ref{table:dictionary2}).\footnote{When viewed from a historical perspective, this equivalence between classical mechanics and an underpinning wave theory is unsurprising, given that Hamilton developed his formulation by analogy with optics \cite{1971Symon}.}

The phase developed along a path from $(t_a,\mathbf{a})$ to $(t_b,\mathbf{b})$ is given by
\begin{equation}\label{eq:goodaction}
\phi = \int_\text{path}  \mathbf{k}d\mathbf{r}-f dt, 
\end{equation}
which can be converted into an integral over time alone by defining the velocity $\dot{\mathbf{r}}\equiv d\mathbf{r}/dt$ and making the other variables parametric functions of time, which gives
\begin{equation}
 \phi = \int_{t_a}^{t_b} \dot{\phi}\, dt =  \int_{t_a}^{t_b}  (\mathbf{k}\mathbf{\dot{r}}-f) dt.
\end{equation}
(In traditional notation this expression reads $S=\int L\, dt = \int (\mathbf{p}\mathbf{\dot{r}}-H)dt$, where $L$ is the Lagrangian.) The extremum condition is enforced by setting $\delta \phi = 0$.  Applying the chain rule then gives
\begin{equation}
\delta \phi = \int_{t_a}^{t_b}  (\delta  \mathbf{k}\,\mathbf{\dot{r}}+\mathbf{k}\,\delta \mathbf{\dot{r}}-\delta f) \, dt.
\end{equation}
Interchanging the order of operations in the  second term gives $\mathbf{k}\,\delta \mathbf{\dot{r}} \, dt= \mathbf{k}\,d(\delta\mathbf{r})$, which can be integrated by parts:
\begin{equation}
\int \mathbf{k}\,d(\delta\mathbf{r}) = \mathbf{k}\,\delta\mathbf{r}\vert_\mathbf{a}^\mathbf{b}-\int \delta\mathbf{r}d\mathbf{k}.
\end{equation}
 We have defined the path to begin at $\mathbf{a}$ and end at $\mathbf{b}$.  While the path varies by $\delta$ between $\mathbf{a}$ and $\mathbf{b}$, at the endpoints $\delta=0$, so the surface term vanishes by construction. Thus we have 
\begin{equation}
\delta \phi = \int_{t_a}^{t_b}  (\delta  \mathbf{k}\,\mathbf{\dot{r}}-\delta \mathbf{r}\mathbf{\dot{k}}\,-\frac{\partial f}{\partial\mathbf{k}} \delta \mathbf{k} - \frac{\partial f}{\partial \mathbf{r}}\delta\mathbf{r}) \, dt,
\end{equation}
where we are taking $f$ to be a function of the position coordinates $\mathbf{r}$, the corresponding spatial frequencies $\mathbf{k}$, and the time $t$.  Collecting terms gives
\begin{equation}
\delta \phi = \int_{t_a}^{t_b}  \left(\delta  \mathbf{k}\, (\mathbf{\dot{r}}-\frac{\partial f}{\partial\mathbf{k}})-\delta \mathbf{r}\,(\mathbf{\dot{k}}+ \frac{\partial f}{\partial \mathbf{r}}) \right)\, dt.
\end{equation}
Because the variations $\delta  \mathbf{r}$ and $\delta  \mathbf{k}$ are independent, their pre-factors must vanish to guarantee $\delta\phi =0$.  Thus we have two relations,
\begin{subequations}\label{eq:goodHamilton}
\begin{align}
\mathbf{\dot{r}}=\frac{d \mathbf{r}}{d t} &=\frac{\partial f}{\partial\mathbf{k}} \qquad \text{and} \label{eq:goodHamilton_groupv}\\
 \mathbf{\dot{k}}=\frac{d \mathbf{k}}{ d t}&=- \frac{\partial f}{\partial \mathbf{r}}, \label{eq:goodHamilton2ndLaw}
\end{align}
\end{subequations}
which are the canonical equations of motion, one for real space and one for reciprocal space, respectively.  Multiplying the first by $h/h$ and the second by $h$ puts them in Hamilton's form:
\begin{subequations}\label{eq:badHamilton}
\begin{align}
\mathbf{\dot{r}}=\frac{d \mathbf{r}}{d t} &=\frac{\partial H}{\partial\mathbf{p}} \qquad \text{and} \\
 \mathbf{\dot{p}}=\frac{d \mathbf{p}}{ d t}&=- \frac{\partial H}{\partial \mathbf{r}},\label{eq:Newton}
\end{align}
\end{subequations}
which define the group velocity and relate the time rate of change of the momentum $\mathbf{p}$ to the gradient in the Hamiltonian $H$, respectively. With the force $\mathbf{F}$ defined as the negative of that gradient, Eq.~\ref{eq:Newton} is Newton's second law. (By definition the potential energy $U$ is the position-dependent portion of $H$, so $ \mathbf{\dot{p}}=- \frac{\partial H}{\partial \mathbf{r}}=- \frac{\partial U}{\partial \mathbf{r}}= - \nabla U\equiv \mathbf{F}$.) 

In traditional language, one says that Newtonian mechanics is equivalent to the statement that the classical physical path extremizes (usually minimizes) the action $S$ \cite{2004Abers}.  In modern language, one says that the classical path extremizes the phase $\phi$, a property intrinsic to waves.  Far from being at odds with wave physics, as is usually implied, classical mechanics follows directly from the most basic and general elements of the wave picture. 

Some comment is required here, since, by writing $\phi$ as an integral (Eq.~\ref{eq:goodaction}) instead of the simple product (Eq.~\ref{eq:phi}), we have passed from a discussion of free waves (Eqs.~\ref{eq:goodeinstein}--\ref{eq:badeinstein}) to one that can accommodate a potential (Eqs.~\ref{eq:goodHamilton}--\ref{eq:badHamilton}).  At the level of fundamental particles, interactions are introduced by postulating the gauge symmetries mentioned earlier and considering one-on-one interactions.  For macroscopic objects the number of waves involved makes this approach impractical, and the sum interaction of many-on-one, or many-on-many is described phenomenologically by adding a potential term that gives $\mathbf{k}$ and $f$, or, equivalently, the Hamiltonian $H$, some dependence on (most commonly) position. Usually this potential term is not obviously Lorentz invariant, as the coordinate dependence is written with respect to a preferred reference frame.  And it may show no sign of its origin as a Lorentz invariant gauge interaction. For instance, the forces that maintain the tension in a pendulum's string, or that push back when a spring is compressed, are at some level electromagnetic, but in these phenomenological descriptions electric charges and electromagnetic fields never appear. Nonetheless the dynamics that follow from the canonical equations of motion (Eqs.~\ref{eq:goodHamilton}--\ref{eq:badHamilton}) can be understood as ultimately arising from the symmetries of $\phi$ under gauge transformations.

Also, for simplicity of exposition, to indicate the real-space position we have been writing $\mathbf{r}$, which in a Cartesian system decomposes to $\mathbf{r}=x \mathbf{\hat{x}}+y \mathbf{\hat{y}} +z \mathbf{\hat{z}}$.  The quantity $\dot{\phi}$, the  equivalent of the Lagrangian in this modern picture (i.e. $L= h \dot{\phi}$), can be written alternatively in terms of generalized coordinates $q_i$, where the $q_i$ are any set of quantities that completely specify the state of the system. Then the wave-equivalent of the generalized momentum canonically conjugate to $q_i$ is $k_i\equiv \partial \dot{\phi}/\partial \dot{q}_i$.  In other words, instead of $\dot{\phi}(\mathbf{r},\dot{\mathbf{r}}, t)$ we can write $\dot{\phi}(q_i,\dot{q}_i, t)$ without invalidating any of the arguments presented above \cite{2002Goldstein,2004Thornton,1976Landau}.  This flexibility, one of the main advantages of the Hamiltonian and Lagrangian formulations relative to the Newtonian formulation of classical mechanics \cite{1971Symon}, is of course unimpaired by the switch to the wave-based description.  (If some $q_i$ depend explicitly on $t$, or the interaction-derived component of $f$ depends on $\dot{q}_i$, then the Hamiltonian is not equal to the total energy, which leads to the preference expressed parenthetically above Eq.~\ref{eq:goodaction} \cite{2004Thornton}.)

The equivalents of energy and momentum conservation, and other consequences of Noether's theorem \cite{2004Hanc,2002Goldstein} (e.g. angular momentum conservation), follow immediately from Eqs.~\ref{eq:goodHamilton}.  The total derivative of $f(\mathbf{r},\mathbf{k},t)$ with respect to time is 
\begin{equation}
\frac{d f}{d t} = \frac{\partial f}{\partial \mathbf{r}} \mathbf{\dot{r}} + \frac{\partial f}{\partial \mathbf{k}} \mathbf{\dot{k}}+\frac{\partial f}{\partial t }.
\end{equation}
The first two terms cancel by Eqs.~\ref{eq:goodHamilton}.  Thus if the frequency $f$ has no explicit time dependence, $d f/dt = 0$ and $f$ is a conserved quantity.  Similarly, by Eq.~\ref{eq:goodHamilton2ndLaw}, if $f$ is independent of some component $q_i$ of $\mathbf{r}$, then $\dot{k_i}=- \partial f/\partial q_i =  0$ and $k_i$, the component of the spatial frequency (or wavevector) conjugate to $q_i$, is conserved.

\begin{table}
\begin{tabular}{llrll}
\multicolumn{2}{c}{traditional} & & \multicolumn{2}{c}{wave} \\
\hline\hline
action&$S$&$=h\times$&$\phi$& phase\\
Lagrangian&$L$&$=h\times$&$\dot{\phi}$& phase time derivative\\
Hamiltonian, energy&$H$, $E$ &$=h\times$&$f$& frequency\\
momentum&$\mathbf{p}$&$=h\times$&$\mathbf{k}$& spatial frequency, \\
& & & & wavevector\\
angular momentum&$\mathbf{J}$&$=h\times$&$\mathbf{J}$& angular momentum\\
force $\mathbf{F}$& $-\frac{\partial H}{\partial \mathbf{r}}$&$=h\times$&$-\frac{\partial f}{\partial \mathbf{r}}$& \\
mass& $m$ & $=\frac{h}{c} \times$&$k_C$& Compton wavenumber
\end{tabular}
\caption{Traditional variables and their wave-based counterparts. In `wave' units $\mathbf{J}$ is taken to be dimensionless. The theory can be completely specified in terms of wave parameters.  To the extent that the concepts on the left are distinct from those on the right,  none of them are necessary.}
\label{table:dictionary2}
\end{table}

\section{Historical perspective}\label{sec:HistoricalPerspective}

Let us compare the traditional and the wave-based developments of classical mechanics.  When Newton wrote $\mathbf{F}= m \mathbf{a}$, he was essentially defining the key concepts of  force $\mathbf{F}$ and mass $m$ in a simultaneous, circular bootstrap.  Operating in complete ignorance of both the space-time connection and the wave properties of matter, Newton had little to work with.  He had to invent his dynamical variables out of nothing, and also to intuit useful relationships between these variables. That his construction survives, after more than three centuries of ceaseless scientific advances, as the practical description of everyday physics is a  testimony to the magnitude of his achievement.  

The Lagrangian and Hamiltonian formulations of classical mechanics, meanwhile, represent another leap in understanding. These formulations make, via the action $S$, the explicit connection between  kinematic variables and dynamic variables that we now recognize as the phase $\phi=-k_\mu x^\mu$, which sits between real space ($x^\mu$) and reciprocal space ($k^\mu$), taking equally from both (Table~\ref{table:dictionary}).  With the Planck-Einstein relation, the de Broglie hypothesis, and the Compton formula,  the dynamical variables are identified as wave parameters and the last connections are finally made.

The wave-based formulation, on the other hand, does not require the incredible improvisations that produced $\mathbf{F}$, $m$, and $S$. The analogous round of definition occurs with the introduction of $f$ and $\mathbf{k}$, but in this case the definitions follow directly from the wave postulate (or even from the definitions of time and distance --- see Appendix~\ref{particles}).  Their meaning is already mathematically precise: these quantities are conjugate to those variables that define the spacetime arena, $t$ and $\mathbf{r}$, and  provide an alternative but equivalent description of the arena's content via the Fourier transform (Appendix~\ref{2pi}). While the wave-based formulation does not remove circularity from the base definitions --- see Section~\ref{sec:waves} and Appendix~\ref{particles} ---  such an achievement is not to be expected \cite{1931Goedel,1936Tarski}. And progress is evident, in that by deriving inertial mass we have removed this important concept from the circle of definitions. The progression from the definitions is also more orderly, proceeding from real space ($x^\mu$) to the boundary ($\phi=-k_\mu x^\mu$), and then to reciprocal space ($k^\mu$), instead of via a giant leap to the dynamical variables followed by a tortuous journey back.

Returning to the discussion of measurement from Section~\ref{sec:waves}, we see that identifying the wave origins of the action $S$ moves this quantity from the realm of the abstract to that of the very concrete, for now we can give it an operational definition. Namely, the action is defined by $S\equiv h \phi$, where $\phi$ is the number of cycles in a wave, relative to some arbitrary origin.  This definition can be compared to that of time $t$, which is the number of ticks of a clock, relative to some arbitrary origin, and position $\mathbf{r}$, which is the number of ruler units in three dimensions, relative to some arbitrary origin.  Unlike time and distance, however, the phase definition is self-contained, in that the counting operation does not rely on the circular (e.g. time-frequency) bootstraps discussed in Section~\ref{sec:waves}. In this sense,  the action $S$, far from being abstract and difficult to understand, is, via its connection to the phase $\phi$, preeminent in the list of well-defined and \emph{measurable} quantities. (Here we are speaking from a perspective that is both `in principle' and strictly classical, in that we are ignoring both problems having to do with the large size of $\phi$ in all but the smallest systems, and those having to do with quantum measurement \cite{1999Espagnat,2019Laloe,2003Bassi,2014Commins,1996Lamb,1966Jammer}.)

Thus the wave-based development requires neither ad hoc definitions nor unmotivated postulates (i.e. $\delta S = 0$).  Three centuries of effort have done more than build a theory that is more accurate and broadly applicable.  They have also revealed underlying principles that make the  theory more coherent and easier to understand.  

 \section{Constants and Concepts}\label{sec:constants}

Traditional pedagogy gives the impression of separate regimes: small systems have wave-like properties, while larger, classical systems do not.  The development presented here shows that these regimes are not so separated.  The subject of classical mechanics is waves: well-localized waves with unmeasurable phases. In other words, classical mechanics treats the limit where the  wave `packets' have phases that are uncountably large and widths that are negligible in comparison to the other length and frequency scales in the problem (see Appendices~\ref{sec:quantum} and \ref{particles}). 

By invoking only three closely-related ideas --- special relativity, waves, and an extremum principle --- we have reproduced the exact (i.e. relativistically correct) dispersion relations, equations of motion, and conservation laws of classical mechanics. Mass has not been inserted as a distinct postulate of the theory, but has emerged as a geometric property of waves in Minkowski spacetime.  We find that the traditional dynamical quantities (Table~\ref{table:dictionary2}) such as action, energy, and momentum are wave properties cloaked in units that disguise these origins.

According to this viewpoint, $c$ and $h$ are not `fundamental' constants, as they are often called, but vestiges of a more primitive (Newtonian) understanding of the underlying physics. Time, distance, energy, momentum, and mass were introduced as distinct aspects of physical theory before the interrelationships were understood.  When, for instance, mass and time seemed unrelated to distance, each was measured in its own distinct units.  Later the relationships were discovered, and constants were introduced to describe the conversions \cite{2004Abers}.

Mass is a particularly interesting case, for the following reason: we experience everyday life in real space.  This fact makes it remarkable, and perhaps even counter-intuitive, that the real-space constraint described by the interval (Eq.~\ref{eq:ssquared}) is not noticeable without scientific equipment, while the reciprocal-space constraint corresponding to mass (Eqs.~\ref{eq:goodeinstein} and \ref{eq:badeinstein}) has obvious real-space manifestations.  Thus the interval $s$ was not measured before it was discovered theoretically --- if it had been known previously, a special unit would have been invented to describe it! But mass has been known since Newton, and its proxy, weight, has been measured since antiquity.  Thus special units were invented to quantify mass before its relationship to other physical quantities was understood.

Despite the misleading language, it is well (if not universally \cite{1985Eisberg,2020Young,2019Inglis}) known that the `fundamental constants', $c$ and $h$, are merely conversion factors set by human convention \cite{2004Abers,1992Taylor,2016Stenger,2012Ralston,2013Ralston,2015Duff}.  The revised International System of Units, or SI, of 2018 recognizes the distinction by purposefully referring  to them as `defining', as opposed to `fundamental', constants \cite{2016Stenger}.  In 1983 and 2018 mortal scientists defined the exact values $c \equiv 299,792,458$~m/s and $h \equiv 6.62607015 \times 10^{-34}$~kg$\cdot$m$^2$/s, respectively \cite{2019Inglis}.  The value of $c$ sets the relationship between the marks on the face of a clock and the marks on a ruler. The value of $h$ sets the relationship between those marks and the marks on a scale, a balance, or a reference mass.  Great pains have been taken to define these constants so as to maintain compatibility with the historical values, and to take full advantage of the best available metrology \cite{2005Borde,2019Inglis}.  But the values human scientists choose to recognize as $c$ and $h$ would not be recognized by alien scientists. In that sense they are not fundamental, but rather the nearly random products of human custom and history. 

Adjusting these constants would be like getting rid of gallons, pounds, and feet in the United States: substantial economic and cultural --- but no physical --- consequences would result.  It is sometimes implied that relativistic and quantum effects are not easily accessed because $c$ is large and $h$ is small, and that, with a god-like power to adjust these constants, we could make our universe appear less classical to our un-aided senses \cite{1964Gamow}. Such statements depend upon non-obvious assumptions, seldom given explicitly, on what constraints are enforced as these constants are adjusted.  

But if the values of these constants are immaterial, why then do we experience the world classically? The answer is that, as complex, sentient beings, we consist of many fundamental particles.  Any being sophisticated enough to do science or metrology must necessarily have a huge number of degrees of freedom.  The question, ``why is Planck's constant so small?''  would be better phrased, ``why is the kilogram so big?'' The answer is that the kilogram consists of a number of atoms that has been (somewhat arbitrarily) chosen to represent a quantity of matter that humans can easily access and manipulate.   Planck's constant is small and the kilogram is big because we consist of very many --- more than Avogadro's number $N_A\equiv 6.02214076 \times 10^{23}$ \cite{2019Inglis} --- fundamental particles. The Planck's constant's small numerical value is \emph{not} ``responsible for the fact that quantum phenomena are not usually observed in our everyday life'' \cite{1964Gamow}.  Rather, quantum phenomena are not usually observed because we consist of many fundamental particles.  The small value for Planck's constant represents a choice that humans have made, and this choice has no impact on any physical phenomenon.
  
For objects consisting of few fundamental particles, large boosts and coherent phase control are achievable with current technology, which allows access to obviously relativistic and  quantum-mechanical phenomena, respectively.  As the number of potentially independent degrees of freedom grows, however, it becomes progressively more difficult to reach the non-classical regimes.  Many particles might be boosted to relativistic velocities via a nuclear explosion, for instance, but such an event is generally hostile to the coherences that maintain a composite object as a unified whole. Quantum phenomena are inaccessible to many-particle objects both because the objects are large in comparison to their wavelengths, and because objects with many degrees of freedom cannot be sufficiently well-isolated from the decohering effects of interactions with the environment \cite{2009Cronin}.  Thus making an observationally less-classical universe would be substantially more involved than simply adjusting $c$ and $h$ --- somehow the observers (e.g. Gamow's Mr. Tompkins) must be constructed from far fewer constituent parts.

While the numerical value of Planck's constant $h$ has no physical significance, we can attribute a significance to its value as the `quantum of action' \cite{2014Messiah}, the very existence of which puzzled Planck himself and  many others since \cite{2014Commins,1966Jammer}.  Here we see that one quantum of action, i.e. $S=h$, corresponds to a phase $\phi=1$~cycle (or $2\pi$ radians), which could hardly be more natural. After all, phase is introduced (Eq.~\ref{eq:phi}) as a countable quantity, and the natural units for counting phase are cycles (see also Appendix~\ref{2pi}). From the wave perspective, the quantization that accompanies the quantum of action also becomes less mysterious. For instance, if energy and frequency are considered to be distinct concepts, one can ask why photons of frequency $f$ necessarily carry energy quantized in increments of $E=hf$.  But does it make sense to ask why photons of frequency $f$ carry frequency $f$?

`Natural' unit systems where $\hbar=c=1$, such as are sometimes employed by particle physicists and cosmologists \cite{1999Jackson,2018ParticleDataGroup,1967Sakurai,1984Halzen,1995Peskin}, bear a resemblance to the wave-based approach presented here. Unfortunately, however, such systems are not employed at the elementary level, and common implementations fail to reap many of the possible  benefits.  For instance, the  electron-volt ($e$V), instead of the second or the meter, is often chosen as the base unit.  This choice is confounding on several levels. As an energy unit,  the $e$V base unit obscures the Fourier transform relationship between physical quantities, such as, for instance, a resonance energy width and the corresponding particle lifetime.  The use of the $e$V is usually restricted to reciprocal space, which further hides the real-space/reciprocal-space connection; adherents to this system rarely quote times or distances in $e$V$^{-1}$ \cite{2018ParticleDataGroup,1984Halzen,1967Sakurai}.  But this particular unit also references electromagnetism.  Earlier we remarked on the logical problems with using energy units to describe the massless photon.  Describing particles such as the neutrino,  the  $\pi^0$, and the neutron, which have no electronic charge, in terms of the electronic charge $e$ is equally inelegant. Energy, momentum, and mass --- or better, the corresponding frequencies --- are general concepts that apply outside the more limited scope of electricity and magnetism, and as such should not be tied to electromagnetic units. 

Another deficiency with $\hbar=c=1$ unit systems, as usually implemented, is not the units \emph{per se}, but rather the vocabulary. These systems maintain energy and momentum as distinct concepts. Just as alien scientists would not recognize our values for $c$ and $h$, so too they might not understand why we distinguish between energy and frequency. The distinction separating these concepts is historical and cultural, not physical. In the wave-based formulation of elementary physics, energy, momentum, and action are superseded.  While not in the same category as phlogiston, caloric, and the luminiferous aether (in that they are not incorrect), these concepts are unnecessary physically and inefficient pedagogically. Their continued use should be discouraged. In the wave-based formulation they never appear, and neither does the conversion factor they require.  The value of Planck's constant, unity or otherwise, is irrelevant. 

As a conversion factor, Planck's constant $h$ is more difficult to defend than the speed of light $c$.  The speed of light connects time and distance.  These concepts are physically and mathematically distinguishable: time has an arrow, while space does not, and time-like coordinates appear with important minus signs relative to their space-like counterparts in the invariants (Appendix~\ref{particles}). While alien scientists would not recognize our value for $c$, they would at least recognize the distinctions we make between time and space.  Planck's constant, on the other hand, connects energy to frequency, momentum to wavevector, and action to phase.  But no physical or mathematical distinction can be made between these quantities, other than the artificial one introduced by Planck's constant itself. 

As has been noted previously \cite{2005Borde}, the existence of `fundamental' constants with dimension indicates that we are calling the same thing by two different names. Some authors see no problem,  emphasizing that the number of dimensional constants is arbitrary \cite{1999Jackson,2004Abers,1971Wichmann}, while others regard dimensional constants as evil \cite{1931Bridgman}. At a minimum having multiple names for one thing can be considered uneconomical, and therefore in violation of a principle that has a long tradition in physics. In his \emph{Principia} Newton himself declares \cite{2016Newton}, as  the first rule of reasoning, that \begin{quote} No more causes of natural things should be admitted than are both true and sufficient to explain their phenomena.\end{quote} As we have argued here, the Newtonian concepts of energy, momentum, and action are not necessary to explain mechanical phenomena --- the more elementary concepts of frequency, spatial frequency, and phase are fully sufficient to construct a complete explanation. Thus Newton's principle of economy dictates that the unnecessary concepts should not be admitted as ``causes''. Adding these concepts to the theory is like adding unnecessary lines to a drawing or unnecessary parts to a machine, to re-purpose a famous analogy \cite{2009Strunk}. Jettisoning energy, momentum, action, and the multitude of conversion formulas involving Planck's constant leaves behind a leaner and cleaner --- but equally capable --- physical theory.

Replacing historical language with wave language gives another perspective on familiar ideas.  On first contact this perspective might seem helpful, or might not.  For instance, the statement \begin{quote}
The energy describes the rate at which the action of an object evolves as a function of time.
\end{quote} 
is quite abstract, and does not evoke a physical picture.  In wave language this statement reads 
\begin{quote} 
The frequency describes the rate at which the phase of a wave evolves as a function of time.\
\end{quote} which is easy to visualize.  On the other hand, making a similar substitution in 
\begin{quote}
The total energy is the sum of the potential energy and the kinetic energy.
\end{quote}
 gives 
\begin{quote}
The total frequency is the sum of the potential frequency and the kinetic frequency.
\end{quote}
which might initially seem bizarre.  That wave language clarifies in some instances, and appears strange in others, can be taken as \emph{prima facie} evidence that the historical vocabulary is hiding potentially valuable ideas. 

For instance, Feynman in his eponymous lectures introduces energy as an abstract-but-conserved quantity. As recorded with his own emphasis, he says \cite{1964Feynman} 
\begin{quote}
It is important to realize that in physics today, we have no knowledge of what energy \emph{is}. 
\end{quote}   
From the traditional perspective this statement is perfectly reasonable --- Newton's ideas offer no underlying foundation for the concept of energy.  However, it is difficult to imagine Feynman saying
\begin{quote}
It is important to realize that in physics today, we have no knowledge of what frequency \emph{is}. 
\end{quote}
These statements are physically equivalent, and thus if the second one is wrong, the first one must be too. Twentieth-century developments not available to Newton provide good insight into what energy is.  To within a `trivial' unit conversion, energy is frequency, where energy/frequency describes the rate at which the action/phase of an object/wave evolves in time. 

The traditional and the modern, wave-based developments can be contrasted as follows.  In the former, the theory is constructed on the makeshift foundation formed with Newton's circular (or nearly so \cite{1964Feynman}) definitions. When the Planck-Einstein, de Broglie, and Compton relations are introduced as 20$^\text{th}$ century developments, they appear rich but unconnected. Without any particular organization or underlying, unifying principle, it is not obvious, for instance, whether the de Broglie relation is better thought of as $\mathbf{p}=h\mathbf{k}$, or $p=h/\lambda$.  Both make the wave connection, but only the former makes the important, additional link to the Fourier conjugate variable.  The wave-based development, on the other hand, has a sturdier foundation, with independent and mathematically precise definitions.  Here the Planck-Einstein, de Broglie, and Compton relations appear as discoveries that connect unit conversions on the three sides of a triangle. While historically important, they are physically trivial.

\section{Scales}\label{sec:scales}

As a practical matter, and to help internalize some sense of the scales involved, it is good to know a few conversion factors.  To an accuracy of better than 1\%, energies and distances are inter-converted  with $h c = 1240\, e$V$\cdot$nm  $= 2 \times 10^{-25}$~J$\cdot$m.  While $e$V$\cdot$nm are usually convenient in condensed matter and atomic physics,  to better fit the typical energy and length scales one might prefer the equivalent units m$e$V$\cdot \mu$m in biological physics, k$e$V$\cdot$pm in electron microscopy, or M$e$V$\cdot$fm in particle physics. 

Masses are inter-converted  with  $h/c=2.21\times 10^{-42}$~kg$\cdot$m and $c^2/h=1.36\times 10^{50}$~Hz/kg. The fantastic size of these numbers reflects the human scale and its complexity.  It is curious to contemplate how such scales are simultaneously both inaccessible (from a direct measurement standpoint) and commonplace (as features of everyday life). We rarely discuss, for example, the Compton wavelengths of macroscopic objects, because they are so small so as to seem unphysical.  The  Planck length $\ell_P \equiv \sqrt{\hbar G/c^3} \sim1.6 \times 10^{-35}$~m, which corresponds to the Planck mass $m_P\equiv \sqrt{\hbar c/G}\sim 22$~$\mu$g, is sometimes thought \cite{1995Garay,1995Peskin,2013Hossenfelder,2014Commins,1973Misner} to represent a minimal length scale,  for instance.  But  a mere kilogram, never mind an astrophysical object, represents a length scale that is far smaller.  Thus the sampling theorem \cite{2000Bracewell} argues against any simple interpretation of the Planck length in terms of a discretization of spacetime.

\section{Conclusion}

Physics is currently taught with a  historical bent that fails to embrace ideas that have been well-established for nearly a century.  Planck's constant is taken to be the signature of quantum mechanical modernity, when it should be viewed as a vestige of `antediluvian' language (the flood here being the development of relativity and quantum mechanics) and a more primitive understanding. The traditional development of classical mechanics is ad hoc, convoluted, and conflates the exact with the approximate. Newtonian ideas and language that are introduced during this development (e.g. energy and momentum) permeate the field from top to bottom. These cultural and linguistic artifacts make physics more difficult than necessary.

An ahistorical, wave-based formulation of `classical' mechanics, on the other hand, is more systematic and economical.  Geometry and symmetry play clear and central roles.  A one-to-one correspondence exists between the kinematic variables $\mathbf{r}$ and $t$ and their dynamical counterparts, $\mathbf{k}$ and $f$.  The equations of motion follow from an extremum principle that is physically motivated.  Mass is not inserted into the theory by hand, but emerges naturally from the geometry of spacetime.  More advanced theories, e.g. quantum mechanics, extend the more elementary material, instead of discrediting it.  In short, the advances of  the 20$^\text{th}$ century  streamline and unify physics from the  17$^\text{th}$ century to the present, so that it is no longer necessary or desirable to so conscientiously retrace the exploratory steps taken by the earliest developers of the theory.

\section{Afterword} \label{sec:afterword}
\subsection{Waves and the Classical Theory}

By assuming waves in this development of `classical' mechanics, we have constructed a theory that is `quantized' from the beginning.  After all, `first quantization' associates the classical variables $E$ and $\mathbf{p}$ with wave parameters $f$ and $\mathbf{k}$, and waves in bound or bounded states can only accommodate quantized values of $f$, $\mathbf{k}$, and $\mathbf{L \equiv r \times k}$.  Thus this reformulation of classical mechanics is hardly classical at all.  

The wave/quantum genesis of this formulation is a strength, not a defect. Our goal is to present physics from a more unified and coherent standpoint.  To the extent that this approach blurs bright lines separating classical from quantum physics, so much the better. The wave-based formulation moves the boundary between these two areas, historically speaking, from 1900 to the mid 1920's, as it explicitly includes (or, more accurately, obviates) the Planck-Einstein, de Broglie, and Compton relations, which are usually considered non-classical.  It also includes the Heisenberg uncertainty relation (1927), which is traditionally considered the archetypical quantum mechanical result: by assuming waves, we have automatically introduced incompatible observables.  Fourier conjugate variables --- $x$ and $k_x$, for instance --- have a restriction on the product of their widths given by 
\begin{equation}\label{goodhup}
\Delta x \Delta k_x \geq 1/4 \pi, 
\end{equation}
which is a purely mathematical result \cite{2000Bracewell,1977Cohen-Tannoudji}.  Thus while the Heisenberg uncertainty relation, \begin{equation}\label{badhup}
\Delta x \, \Delta p_x \geq \hbar/2,
\end{equation}  
might seem mysterious in traditional units, it has both a solid mathematical foundation and a simple, intuitive explanation in wave units.  One does not need a physicist's mathematical training to understand that it is not possible to simultaneously measure, for example, time and frequency to arbitrary precision.

\subsection{Choice of Wave Equation}
As the attentive reader might have noticed, the motivation accompanying the introduction of $\phi$ in Section~\ref{sec:waves} has a shortcoming. The generic wave equation (Eq.~\ref{eq:genericwave}), which is invoked but not actually used, is not invariant under a Lorentz transformation unless the phase velocity $v=c$, the speed of light.  Thus it cannot be applied to massive bodies and a more general wave equation must be sought. 

Unlike Eq.~\ref{eq:genericwave}, the Klein-Gordon equation
\begin{equation}\label{eq:kleingordon}
-\frac{1}{c^2}\frac{\partial^2 u}{\partial t^2} + \nabla^2 u = (2 \pi k_C)^2 u,\tag{\thewavecounter b}
\end{equation} provides  a natural fit with Eq.~\ref{eq:goodeinstein}. We decline to take this step in the first pass because the choice of wave equation is unimportant for the development of classical mechanics, and the generic wave Eq.~\ref{eq:genericwave}, unlike the Klein-Gordon Eq.~\ref{eq:kleingordon}, is familiar from a variety of classical problems (e.g. sound, vibrating strings, electromagnetic waves).  However, in a second pass Eq.~\ref{eq:kleingordon} is preferred for its ability to accommodate mass while preserving Lorentz invariance. While seen less frequently, Eq.~\ref{eq:kleingordon} can be employed in classical contexts --- for instance, to describe the braced string \cite{2011Gravel,1968Crawford}.

Because it fails to directly address (or even identify) the underlying wave equation and associated concepts such as superposition and dispersion, the wave-based formulation of classical  mechanics is manifestly incomplete.  Rectifying this shortcoming leads to developing the full quantum theory.  This development dramatically expands the number of postulates (adding e.g. states are represented by vectors in a Hilbert space, etc.) \cite{1977Cohen-Tannoudji,2018Griffiths,1999Espagnat,2018Neumann,2003Bassi,2014Commins,1968Saxon}, which is an indication that this next level is not as well understood. 

\appendix

\section{Fourier conjugate variable choice}\label{2pi}

While employed without comment previously, the choice of the Fourier conjugate variable $f$ for frequency over its radial counterpart $\omega\equiv 2 \pi f $ is deliberate, and in keeping with the goal of economy. Introducing radial frequencies does not obviously reduce the number of $2 \pi$'s, but it increases the number of variables in use (unless one decides to forgo $f$ entirely).  The additional variable $\omega$ adds a new way to write any formula containing $f$, but without adding any new idea or perspective.  Thus it increases algebraic clutter for an at-best marginal decrease in the number of $2\pi$'s.  

The ordinary frequency $f$ is preferred over the angular frequency $\omega$ for two reasons.  First, it  puts the Fourier transforms in their symmetric, unitary form 
\begin{subequations}\label{eq:timeFT} 
\begin{align}
 \bar{u}(f) &= \int_{-\infty}^{\infty} u(t)e^{2\pi i f t} dt\\
  u(t) &= \int_{-\infty}^{\infty} \bar{u}(f)e^{-2\pi i  f t } df.
\end{align}
\end{subequations}
This form simplifies the interpretation of many transform pairs.  For instance,  the Dirac $\delta$-function and unity are Fourier transforms of one another (i.e. for $u(t)=\delta(t)$, $\bar{u}(f)=1$, and for $\bar{u}(f)=\delta(f)$, $u(t)=1$), without factors of $2\pi$ or $\sqrt{2\pi}$. (As Fourier conjugate variables $E$ and $\mathbf{p}$ are worse than either choice of the frequency variables, for then normalization factors of Planck's constant $h$ or Dirac's reduced-Planck's-constant $\hbar$ also appear.) Second and most importantly, $f$ preserves the simplest and most symmetrical reciprocal relationship between the period $\tau \equiv 1/f$ and the temporal frequency $f$.  

The wavevector, or spatial frequency, $\mathbf{k}$ is defined analogously, in that the wavelength $\lambda \equiv 1/|\mathbf{k}|$ is like the period $\tau \equiv 1/f$.  This wavevector convention differs from the usual physicist's convention by a factor of $2 \pi$.  However, it is a standard choice in mathematics \cite{2003Stein,1989Folland}, engineering \cite{2000Bracewell}, crystallography \cite{1996Schwarzenbach}, and in X-ray \cite{1969Ewald,2019Jacobsen} and electron microscopy \cite{2008Reimer,1995Cowley}, where the Fourier conversion between real space (imaging) and reciprocal space (diffraction) is a frequently-performed, often push-button operation. The convenience advantage in  being able to go back and forth, without having to wrestle with $2\pi$'s, between wavevectors and real-space distances is difficult to overstate. Besides giving the simplest (reciprocal) relationship between wavelength $\lambda$ and wavevector $\mathbf{k}$, the microscopist's convention also streamlines concepts such as the density of states --- compare, for example the formulas $(L/2\pi)^D$ and $L^D$ for the density of states in $\mathbf{k}$-space for a  $D$-dimensional volume $L^D$.

From the perspective of the phase $\phi$ (a `countable' quantity --- see discussion after Eq.~\ref{eq:phi}) as well, ordinary frequency is preferred over radial frequency.  Counting cycles in a waveform can often be accomplished by inspection.  Counting radians is not so easy.

\section{Sign conventions}
We choose a metric $(+---)$ that gives $x^\mu x_\mu >0$ for timelike separations and $p^\mu p_\mu \propto k^\mu k_\mu \propto m^2 >0$ for normal objects. For consistency with the standard quantum mechanics conventions for $H= h f$ and $\mathbf{p}= h\mathbf{k}$ (first two lines of Table~\ref{table:qm}) we choose $u(x^\mu) = e^{2\pi i (\mathbf{k' r} - f't)} = e^{-2\pi i k' \cdot x} = e^{2 \pi i \phi}$ to describe a plane wave with $\mathbf{k'}>0$ and $f'>0$.  Because of the hyperbolic metric, we must have (compare the signs in Eqs.~\ref{eq:timeFT}),
\begin{subequations}\label{eq:spaceFT} 
\begin{align}
 \bar{u}(\mathbf{k}) &= \iiint_{-\infty}^{\infty} u(\mathbf{r})e^{-2\pi i \mathbf{k\cdot r}} d^3\mathbf{r}\\
  u(\mathbf{r}) &= \iiint_{-\infty}^{\infty} \bar{u}(\mathbf{k})e^{2\pi i  \mathbf{k\cdot r}} d^3\mathbf{k}.
\end{align}
\end{subequations}
Equations \ref{eq:timeFT}--\ref{eq:spaceFT} can be summarized
\begin{subequations}\label{eq:spacetimeFT} 
\begin{align}
 \bar{u}(k^\nu) &= \iiiint_{-\infty}^{\infty} u(x^\nu)e^{2\pi i k \cdot x} d^4x\\
  u(x^\nu) &= \iiiint_{-\infty}^{\infty} \bar{u}(k^\nu)e^{-2\pi i  k \cdot x} d^4k,
\end{align}
\end{subequations}
where $k \cdot x \equiv k_\mu x^\mu$. In Dirac's notation $\bra{k}x\rangle =e^{2\pi i k \cdot x}$.

We choose $\phi = - k_\mu x^\mu$ (Eq.~\ref{eq:phi}) with the leading minus sign so that the conversion between wave units and traditional units for the action $S=h \phi$ will be like those for the energy  $H = h f$ and the momentum $\mathbf{p}= h\mathbf{k}$.  The sign of the action/phase is purely conventional, since the trajectories are independent of whether the extrema is a minimum or a maximum (Section~\ref{sec:stationary}).  Historically the choice has been to have $dS/dt \equiv L  \equiv \mathbf{p}\mathbf{\dot{r}}-H$ and not $dS/dt \equiv L \equiv H-\mathbf{p}\mathbf{\dot{r}}$, which would have been better for the uniformity of our sign conventions. The choice of the sign defining the Lagrangian $L$, and thus the action, in this Legendre transformation was likely made because $\mathbf{p}\mathbf{\dot{r}}>H$ for a free particle if its rest energy is ignored, and the concept of rest energy was unknown when this convention was established.

\section{Quantum mechanics and its classical limit}\label{sec:quantum}

Quantum mechanics is purported to present a picture that is wildly distinct from that of classical mechanics \cite{2018Griffiths}. Classical mechanics is the intuitive, every-day world of massive bodies, while quantum mechanics involves spooky, hard-to-understand waves \cite{1985Mermin}.  Our purpose in this article is to de-emphasize that distinction. Classical mechanics is also a wave theory.  It happens that classical mechanics was invented before its wave underpinnings were understood, but we can understand both it and subsequent developments better by exploiting those same developments.  Emphasizing the differences between classical and quantum physics has few pedagogical advantages, and understanding quantum mechanics is much easier if one has a classical intuition that is already wave-based.

Quantum mechanics has some real mysteries: entanglement, locality, and the measurement problem, for instance. However, other aspects of quantum mechanics that can seem problematic are not, such as the uncertainty principle and incompatible observables.  It might seem deep and mysterious that one cannot simultaneously measure position and momentum \cite{2018Griffiths}.  But, as discussed in Section~\ref{sec:afterword},  it is comparatively obvious that one cannot simultaneously measure position and spatial frequency. 

Using modern, wave-based units (i.e. dispensing with Planck's constant) will improve the clarity of any presentation of quantum mechanics, which is manifestly a wave theory. But given that many excellent references \cite{1967Dirac,1981Landau,2010Feynman,2018Griffiths,2014Commins,1964Gamow,2014Messiah,2002Goldstein,1977Cohen-Tannoudji,1991Cheng,2004Abers,2017Sakurai,1968Saxon} contain confusing statements about the role of Plank's constant, particularly in regards to the classical limit, a few additional words are in order. 

Taking $\hbar\rightarrow 0$ is often cited as a means of reducing a quantum result to the classical limit \cite{1967Dirac,1981Landau,2010Feynman,2018Griffiths,2014Messiah,2002Goldstein,1991Cheng,2004Abers,2017Sakurai,1999Jackson,1968Saxon,1966Jammer}.  As we have shown, Planck's constant $h=2 \pi \hbar$ is not a necessary element of physical theory: if wave-based units are used from the beginning, this conversion factor need never enter the discussion.  And even if we do decide to use traditional units, then adjusting the size of Planck's constant is equivalent to changing the definition of the kilogram relative to the meter and the second (Section~\ref{sec:constants}). Such re-definition of the agreed-upon standard unit of mass ought not to have physical effects.  Thus, according to the perspective presented here, the value of Planck's constant --- physically irrelevant, and set by custom --- should have no bearing on the classical limit. 

Another way to recognize this problem is to note that some authors set $\hbar = 1$, others state that the classical limit occurs as $\hbar \rightarrow 0$, and some do both \cite{1991Cheng,2004Abers,1999Jackson}.  Taking both statements at face value gives the nonsensical claim ``$\hbar = 1 \rightarrow 0$ in the classical limit'', indicating that one of these assignments is of questionable generality.

Some equations support the $\hbar\rightarrow 0$ sophistry, but it is not true generally and indiscriminate application can lead to nonsense.  For instance, applying this `rule' to the canonical commutation relation 
\begin{equation}\label{eq:canonical}
[x,p_x]= i \hbar
\end{equation}
and its corresponding uncertainty relation $\Delta x \Delta p_x \geq \hbar/2$ (Eq.~\ref{badhup}) gives results that are sensible, if not revealing.  These relations imply that position and momentum are classically compatible, as indeed they are: both variables can be simultaneously known to arbitrary precision within a classical framework.

But the $\hbar\rightarrow 0$ rule gives results that are difficult to defend in  what might be the very next topic \cite{1964Feynman}.  Applying the canonical commutation relation \ref{eq:canonical} to the angular momentum operators $\mathbf{L} \equiv \mathbf{r} \times \mathbf{p}$ gives the commutator
\begin{equation}
[L_x, L_y] = i \hbar L_z.
\end{equation}
The angular momentum operators are the  generators of rotations \cite{1977Cohen-Tannoudji,2002Goldstein}. Their non-zero commutator says that rotations in three dimensions do not commute, as is true both quantum-mechanically and classically. 

Not only does   $\hbar\rightarrow 0$ give a geometrically untenable result here (namely that rotations in three dimensions \emph{do} commute)\cite{2012Ralston}, it also muddies an opportunity to examine a general topic of wide applicability. Non-commuting rotations give rise to `geometric' phases, which appear in the theory at the same level as the `dynamical' phase $\phi$.  (In other words, the `geometric' and the `dynamical' phases are additive.)  While elementary enough to be demonstrated with just an arm and a thumb \cite{1989Chiao}, geometric phases have classical, quantum mechanical, and gauge theoretic relevance. Examples of topics elucidated with geometric phases include the Foucault pendulum and cyclotron motion, the Aharonov–Bohm and Aharonov–Casher effects, Wess-Zumino terms, their anomalies, and fractional statistics \cite{1989Shapere,2019Cohen}. 

In wave-based units the angular momentum operators are dimensionless, and the $\hbar\rightarrow 0$ problem with their commutator does not arise.  There is never any implication that rotations in three dimensions commute. A classical limit exists in the sense that, with a sufficiently large angular momentum, its three orthogonal components can be simultaneously determined.  This limit corresponds to a failure to track the phase at the single cycle level (see below).  

The case of rotations, with its geometric roots and applications throughout classical and quantum physics, is thus clearly important, and here the $\hbar \rightarrow 0$ prescription fails.  The prescription can be patched by adding a special translation for relating the classical Poisson bracket to the quantum mechanical commutator.  The translation $[u,v] \rightarrow (1/i \hbar) (uv-vu)$ gives the correspondence between classical functions on the left and quantum operators on the right \cite{2002Goldstein,2017Sakurai,1968Schiff}, but the appearance of $\hbar$ in the denominator of the patch just emphasizes that $\hbar$ must drop out of the problem.

Thus the blanket prescription $\hbar \rightarrow 0$ is not a reliable route to the classical limit. But we recognize real differences between the quantum and the classical regimes, and it is helpful to have a prescription for passing from the former to the latter.  How then is the classical limit found without Planck's constant, and to what extent does the $\hbar\rightarrow 0$ route to the classical limit make sense?

Examining the wave-based form of the canonical commutation relation, 
\begin{equation}
[x,k_x]=i/2\pi,
\end{equation}
and the associated uncertainty relation, 
\begin{equation*}\tag{\ref{goodhup}}
\Delta x \Delta k_x \geq 1/4\pi,
\end{equation*}
provides a more constructive perspective.  The classical limit does not correspond to taking the right-hand sides to zero, for $i/2\pi \rightarrow 0$ and $1/4\pi \rightarrow 0$ do not make sense. In fact, in the classical limit the product on the left-hand side of the uncertainty relation Eq.~\ref{goodhup} does not even need to be large, and often it is not (see Appendix~\ref{particles}). Coherent and displaced states of the `quantum' harmonic oscillator, which give the equality in Eq.~\ref{goodhup} and exhibit classical, oscillatory position and momentum expectation values \cite{2018Griffiths,1977Cohen-Tannoudji,1968Saxon,1968Schiff}, provide well-known examples of this general phenomenon.

In the classical limit trajectories are well-defined, so $x \gg \Delta x$ and $k_x \gg \Delta k_x$ \cite{1977Cohen-Tannoudji,2014Messiah}. If $k_x \gg \Delta k_x$ we can take $\Delta k_x/k_x \simeq \Delta \lambda/\lambda$, which gives $\lambda\gg \Delta \lambda$. Thus, the uncertainty relation (Eq.~\ref{goodhup}) can be written $ \Delta x \Delta k_x \simeq \Delta x \Delta \lambda/\lambda^2\geq 1/4 \pi$. Satisfying  the requirements for classical trajectories and the uncertainty relation simultaneously thus requires that 
\begin{equation}\label{inequalities}
\MyFourTerm,
\end{equation}
which holds even in the limiting case (equality) of the uncertainty relation. Note a subtlety here ---  we expect the uncertainties to be small, not large, in the classical limit. However, one can legitimately consider $\Delta x \rightarrow \infty$ or $\Delta x \rightarrow 0$, depending on the basis for the comparison (which points out the hazards of limits in dimensioned quantities).  In other words, with two (or more) other length scales in the problem, the uncertainty $\Delta x$ can be both large and small. It is large in comparison to the wavelength $\lambda$.  And it is small in comparison to $x$, a classical dimension in the problem (not necessarily the position coordinate, which can be set to zero by choice of origin).
 
Simple estimates of $x$ and $\lambda$ put  macroscopic objects properly in the classical limit.\footnote{Exploring the quantum/classical boundary with nanoscopic objects is an active area of research \cite{2009Cronin}. Quantum interference with $\Delta x \gg \ell$ has been demonstrated using 2000-atom molecules (e.g. C$_{707}$H$_{260}$F$_{908}$N$_{16}$S$_{53}$Zn$_{4}$) with masses of $2.5\times 10^4$~Da and size-to-wavelength ratios $\ell/\lambda$ of $10^5$ \cite{2019Fein}. (One dalton, or unified atomic mass unit, is equivalent to approximately $0.75$~fm$^{-1}$.) Achieving interference with masses exceeding $10^5$~Da looks feasible with current technology \cite{2019Kialka}.}  The atomic size scale is the Bohr radius $a_0=  \hbar/(\alpha m_e c)$, where $\alpha$ is the fine-structure constant and $m_e$ is the electron mass, so an object's size $\ell\sim x\sim a_0 N^{1/3}$, where $N$ is the number of atoms in the object. For  $\lambda$, a conservative (i.e. large $\lambda$) estimate corresponds to an object that has no apparent motion. Ascribing the object's motion  to its thermal energy $k_B T$ alone gives $p^2/M = k^2/k_C\sim k_B T$.  Then $\lambda \sim h/ \sqrt{N A m_n k_B T}$, where the object's mass $M$ comes from its $N$ atoms having $A$ nucleons of mass $m_n$. Comparing $N^{1/3}$ to $N^{-1/2}$, we see that, for a macroscopic quantity $N$ (e.g. $N_A$, Avogadro's number), $\ell\gg \lambda$ independent of any other details, and that the range where $\Delta x $ can simultaneously satisfy the inequalities \ref{inequalities} spans many orders-of-magnitude.

In passing we note that applying $\hbar \rightarrow 0$ to the first argument of the previous paragraph gives the conclusion that solids do not exist in the classical limit. This statement is true in the sense that `classical' atoms are unstable to Coulombic collapse, and it leaves classical mechanics bereft of practically all of its usual and fruitful fields of application. From a wave perspective the stability of atoms is unsurprising, and the classical limit corresponds to taking the dimensionless $N\rightarrow \infty$. 

Many authors draw an analogy between the route from physical (i.e. wave) optics to geometrical optics given by $\lambda \rightarrow 0$, and the route from quantum mechanics to classical mechanics given by $\hbar\rightarrow 0$ \cite{2014Messiah,1977Cohen-Tannoudji,1981Landau,2017Sakurai}.  In traditional units this comparison is fairly called an analogy, but in wave units the analogy is so apt that it is better viewed as an equivalence. (The wave picture tightens many analogies between optics and mechanics, such as that between a refractive index and a potential energy \cite{2009Cronin}.) Just as one passes from physical optics to geometric optics by taking $\lambda \rightarrow 0$ (or $f \rightarrow \infty$), so one passes from quantum mechanics to classical mechanics. 

The $\hbar\rightarrow 0$ approach to the classical regime for massive particles (this caveat is important --- see next paragraph) is thus better understood as concerning the limit where the frequencies are large in comparison to the other (inverse) time or length scales in the problem. In other words, in the Planck-Einstein relation $E=hf$ or the de Broglie hypothesis $\mathbf{p}=h \mathbf{k}$, the classical variable ($E$ or $\mathbf{p}$) is \emph{held constant implicitly} while Planck's constant is taken to zero, which means that the frequencies are taken to infinity.  In this limit the phase $\phi$ is not being  tracked at the single cycle level, which is equivalent to the action $S$ not being tracked with a precision of $h$ \cite{2003Hanc}.  

The $\hbar\rightarrow 0$ prescription does hold one remarkable advantage for the expert practitioner. By judiciously choosing  the `classical' variable to hold fixed, one can successfully identify the classical limit for both the massive and the massless cases.  Massive particles have an associated energy $E$ classically, and thus taking $\hbar\rightarrow 0$ gives $f\rightarrow \infty$ as just described.  A massless (e.g. electromagnetic) wave, on the other hand, is classically identified by its frequency.  Here $\hbar\rightarrow 0$ gives $E\rightarrow 0$, which indeed is the classical limit.  For instance, the quantum result for a blackbody's energy per mode, $h f/(\exp[hf/k_BT]-1)=E/[\exp(E/k_BT)-1]$, reduces to $k_BT$ in the classical limit. 

Turning these arguments around reminds us that the `quantum vs. classical' limit is not the same as the `wave vs. object/particle' limit.  Traditional language gives a confusing, crossover behavior, namely that, in the quantum limit, classical particles adopt wave properties, while classical waves adopt particle properties \cite{1999Espagnat}. For example, consider a double-slit experiment performed with electrons or low-intensity electromagnetic waves. The former show quantum behavior by going through both slits, while the latter show quantum behavior by hitting the screen in localized quanta, i.e. photons. In wave-based language the perspective is simpler: $\lambda\rightarrow 0$ (or $f \rightarrow \infty$) produces particle behavior and  $\lambda\rightarrow \infty$ (or $f \rightarrow 0$) produces wave behavior always. In traditional language the correspondence between these limits and the quantum/classical limit depends on whether the objects under consideration are massive or massless.

\begin{table}
\begin{tabular}{rclrcl}
\multicolumn{3}{c}{traditional} & \multicolumn{3}{c}{wave} \\
\hline
$H$&=&$i \hbar \frac{\partial}{\partial t}$ & $f$ &=& $\frac{i}{2 \pi} \frac{\partial}{\partial t}$\\
$\mathbf{p}$&=&$\frac{\hbar}{i} \nabla$ & $\mathbf{k}$ &=& $ \frac{1}{2\pi i} \nabla$\\
$[x,p_x]$&=&$i \hbar $ & $[x,k_x]$ &=& $ i/2\pi$\\
$\Delta x\Delta p_x$&$\geq$& $\hbar/2$ & $\Delta x \Delta k_x$&$\geq$& $1/4\pi$\\
$\mathbf{L} $&=&$ \mathbf{r} \times \mathbf{p}$ & $\mathbf{L} $&=&$ \mathbf{r} \times \mathbf{k}$\\
$[L_x,L_y]$&=&$i \hbar L_z $ & $[L_x, L_y]$ &=& $ \frac{i}{2\pi} L_z$\\
$U(t')$&=&$e^{i H t'/\hbar}$ &  &=& $e^{i 2 \pi f t'}$\\
$T(\mathbf{r'})$&=&$e^{-i \mathbf{p}\cdot \mathbf{r'}/\hbar}$ & &=&$e^{-i 2\pi \mathbf{k}\cdot\mathbf{r'}}$\\
$B(\mathbf{p'})$&=&$e^{i \mathbf{p'}\cdot \mathbf{r}/\hbar}$ &$B(\mathbf{k'})$&=&$e^{i 2\pi \mathbf{k'}\cdot\mathbf{r}}$\\
$R(\mathbf{\theta'})$&=&$e^{-i \mathbf{L}\cdot \mathbf{\theta'}/\hbar}$ & &=&$e^{-i 2\pi \mathbf{L}\cdot\mathbf{\theta'}}$\\
$\frac{d A}{dt}$&=&$\frac{i}{\hbar}[H,A]+\frac{\partial A}{\partial t}$ &  &=& $ 2 \pi i[f,A]+\frac{\partial A}{\partial t}$
\end{tabular}
\caption{Standard quantum mechanical relations written in traditional \cite{2018Griffiths,1977Cohen-Tannoudji,2002Styer} and wave-based units, i.e. with and without the reduced Planck constant $\hbar$.   Some important quantum mechanical operators, such as the time evolution, translation, boost, and rotation operators, are functions of $1/\hbar$ in traditional units, and thus become meaningless in the limit $\hbar \rightarrow 0$.}
\label{table:qm}
\end{table}

Table~\ref{table:qm} gives a list of standard quantum mechanical formulas in traditional units side-by-side with their wave-unit counterparts.  The equations with $\hbar$ in the denominator become meaningless if one approaches the `classical limit' by taking $\hbar\rightarrow 0$.  In this way the $\hbar\rightarrow 0$ prescription fails for an operation such as, for instance,  translation, which seems sufficiently mundane that it ought to have a well-defined classical limit. The equations with $\hbar$ in the denominator also illustrate a general principle: if a physical quantity is being tracked in classical units, it must be converted out of that unit system before it can be used to actually do anything. In this sense the wave-based unit system is `natural' and the traditional system, with its energies and momenta, is not. The time-evolution, translation, boost, and rotation operators all illustrate this principle.

\section{The quantum limit of classical mechanics}\label{sec:classical}
Traditional classical mechanics gives no prescription for finding the quantum limit, but with the wave-based formulation we can create one by reversing the arguments of Appendix~\ref{sec:quantum}. The situation is not exactly symmetric, in that we cannot reproduce a prediction of the exact theory by taking a limit in the approximate theory, but we can at least identify where the approximate theory breaks down. Just as the classical limit of quantum mechanics is found by taking $\lambda$ to be small, so the quantum limit of classical mechanics is found by taking $\lambda$ to be \emph{not} small. 

When we interpret such limits involving a dimensioned quantity, the existence of another, comparison scale is always implied.  Quantum mechanics is widely understood to be the theory of small systems.  Here `small' means that other length scales in the problem are comparable to the wavelength $\lambda$.   At this point the crude, approximate description generated by considering the stationary behavior of the phase (i.e. the action) alone is no longer sufficiently accurate, and the specifics of the wave equation must be considered.  Expressing these ideas in traditional language, one expects the classical theory to break down when the energy $E$ and momentum $\mathbf{p}$ become small (giving small frequencies and large wavelengths).  However, while wave units provide a comparison scale --- namely the size of the system ---  traditional units do not provide comparison scales for the dimensioned quantities  $E$ and $\mathbf{p}$. Viewed from a purely Newtonian perspective, the energy and momentum can be made arbitrarily small without entering a new regime that might be governed by qualitatively different physical laws. Thus, without the advantages of the wave-based perspective, traditional classical mechanics  gives no hint that it is incomplete --- hence the surprise accompanying the quantum revolution of the early 20$^\text{th}$ century.

\section{Glossary}\label{particles}

To avoid slowing our development, in a few cases we have adopted the practice borrowing `physics words', which ideally have precise meanings, from common English without pausing to define them. Because of the bootstrapping problem, it may be impossible to articulate definitions that are wholly satisfactory \cite{1971Symon}, so this practice is understandably common. Most authors adopt it implicitly. Others warn that they are capitulating with a few words such as ``we can intuitively sense the meaning of mass'' \cite{1964Feynman}, or ``length, time, and mass are concepts normally already understood'' \cite{2004Thornton}, or ``fundamental physical concepts, such as space, time, simultaneity, mass, and force...will not be analyzed critically here; rather, they will be assumed as undefined terms whose meanings are familiar to the reader'' \cite{2002Goldstein}. Newton himself writes, ``I do not define time, space, place, and motion, as being well known to all'' \cite{2002Hawking}.

Here we have carefully defined space and time in terms of their respective clock frequencies, and vice versa (Section~\ref{sec:waves}).  We have argued that the dynamical quantities energy and momentum are just frequencies multiplied by a dimensionful constant that is a historical legacy. And we have defined mass as the corresponding hypotenuse of a hyperbolic right triangle that has temporal frequency and spatial frequency as its other sides. In so doing we have defined most of the concepts that appear in Tables~\ref{table:dictionary} and \ref{table:dictionary2}, which together cover classical mechanics.

However, the terms appearing in the bottom row of Table~\ref{table:dictionary} merit further discussion. As stated earlier, an event is a position $x_\mu$ in real space.  From the structure of Table~\ref{table:dictionary} then we deduce that a wave is a position $k_\mu$ in reciprocal space, and an object is that same position, given instead in historical units.  This usage of the term `wave' is in keeping with the common usage of the term: a wave has a well-defined frequency and spatial frequency.

However, as the Heisenberg uncertainty principle (Eq.~\ref{goodhup}) explains, a wave/object exactly located in reciprocal space is completely delocalized in real space. Temporal delocalization is familiar --- we see no problem with an object persisting for a long time --- but spatial delocalization is in contradiction with everyday experience.  Thus, while waves corresponding to classical objects must have well-defined frequencies/momenta, these cannot be exact: they must have some spread.  The waves/objects under discussion (i.e. those in the classical limit) must take the form of wave packets that are localized in both real- and reciprocal-space.  

Students familiar with Fourier transforms might protest that, since a small width in one space implies a large width in the reciprocal space ($\Delta x \propto 1/\Delta k$), it is impossible to be well-localized in both.  The response to this objection is that, in this context, we are not determining size by comparing $\Delta x$ to $ 1/\Delta k$.  Rather, here by well-localized we mean $x \gg \Delta x$ and $k \gg \Delta k$.  In other words, the magnitudes of the classical coordinates in real- and reciprocal-space are large compared to the widths of these wave packets \cite{1977Cohen-Tannoudji,2014Messiah}.

By a change of origin the coordinates of any one location, either in real or reciprocal space, can be made to be zero.  However, this condition cannot be simultaneously achieved for all the constituents of an object consisting of many.  In real space the classicality condition can be satisfied by taking the object's classical length scale `$x$' to be its spatial dimension $\ell$ (i.e. its size), and that scale to be much larger than the real-space packet's width.  

But the reciprocal-space case requires a more nuanced argument.  For instance, one might wonder why an object --- say a billiard ball --- at rest is not delocalized.  After all, in the frame where the billiard ball is motionless, one might think that its momentum is both small ($\mathbf{p}\sim 0$) and well-specified ($\Delta \mathbf{p} \sim 0$), which would imply that $\lambda$ and $\Delta \mathbf{r}$ are both large (via the de Broglie and uncertainty relations respectively). This conundrum is resolved by noting that an ``object's rest frame'' only exists in the classical approximation.  One can be in, or transform to, a frame where an object's center of mass is at rest.  However, for a multi-particle object this point is only a mathematical construction relative to which the object's constituent parts are generally in motion. 

Consider, for example, the object as a collection of $N$ coupled harmonic oscillators, which is a reasonable first approximation of a solid. Any collection of coupled harmonic oscillators of natural frequency $f_0$ will produce normal modes with frequencies both above and below $f_0$ \cite{2004Thornton}, and the frequencies of the lowest-frequency normal modes scale like $f_0/N$ \cite{2005KittelIntroduction}.  At any finite temperature these low-frequency modes are thermally excited, and thus the object consists of, say, halves in thermal motion moving about the center of mass. The equipartition theorem then implies that these halves have thermal momenta such that $\lambda \sim h/\sqrt{M k_B T}$, a small length for a macroscopic mass $M$ (Appendix~\ref{sec:quantum}).  Thus the halves are localizable, and so must be the object itself.  This argument applies to all three spatial directions separately, in that one cannot transform, by either rotation or boost, to a frame where a classical (i.e. many-particle, finite-temperature) object is delocalized in any spatial dimension. Thus the $\MyFourTerm$ condition discussed in Appendix~\ref{sec:quantum} is achieved.

We have now introduced enough ideas from both classical and quantum physics that we can examine the definition and usage of these three key words: particle, wave, and object. (Other terms, such as `corpuscle', `thing', `entity', and `body', are sometimes used synonymously and interchangeably with one or more of the three terms we have chosen.)  The first term in particular is used with a bewildering array of meanings, some quite contradictory --- even compared to other physics concepts, the definition of  `particle' is particularly elastic.  Standard classical mechanics textbooks \cite{2020Young,1976Landau,2004Thornton,2014Kleppner,2002Goldstein,1971Symon} use `particle'  to denote the concept of an `object' (as used here) or a `body' (in Newton's language) \cite{2016Newton}. Some emphasize a negligible spatial extent, defining a particle to be a point-like object \cite{1971Omnes,1976Landau,1971Symon}. But in more advanced textbooks usage inconsistent with these definitions is common.  For instance, the wavefunction $\psi \sim e^{i (\mathbf{p}\cdot \mathbf{x} -  E t)/\hbar}$ might be given to describe a free particle of energy $E$ and momentum $\mathbf{p}$ \cite{1984Halzen,1968Saxon,1999Espagnat}. This state is completely delocalized in real space, and thus could not be less point-like. Unfortunately, terminology that provides an explicit warning (e.g. use of `wavicle' \cite{1928Eddington}) has not been widely adopted.

Identifying, for instance, a particle such as an electron or the muon as a ``point particle'' \cite{2000Harrison,2004Abers,2010Griffiths,1985Eisberg,1999Jackson,1984Halzen} is particularly misleading. An electron can be never be localized to a point, and is often delocalized over macroscopic dimensions (as in an electron microscope, for instance).  Under certain circumstances (e.g. only the center of mass motion is of interest, or the point of observation is distant from an object with spherical symmetry) we can treat an extended object/wave as a mathematical point, but under no circumstances is it proper to think of the ``point particle'' as anything other than a convenient approximation. A diffracted electron does not hit a phosphor screen at a point.  Rather, it hits the screen in a locale that might be identified with optical ($\simeq 1$~$\mu$m) or even atomic ($\simeq 0.1$~nm) spatial resolution, but not better.  The collision with the screen produces a measurement of the electron's position, but this measurement does not `convert' the wave-like electron into a point-like particle, even for an instant \cite{2004Abers}.  The electron is still wave-like, and after it collides with the screen it occupies wave-like states in the screen. Size is not an intrinsic property of the electron, and thus one cannot ascribe it any permanent value, zero or otherwise. The size an electron manifests depends on how aggressively it is localized or probed.  In a deep potential well, or subject to a collision with a high energy particle, an electron can be well-localized (down to the length scale corresponding to its Compton wavelength, at which point pair production prohibits further progress).  A free electron, or an electron in a clean crystal, can be delocalized over macroscopic length scales.

Having identified a few hazards, we here collect some of our previous definitions and attempt explicit definitions of some of the terms remaining: 
\begin{enumerate}
\item Time $t$ is measured by counting clock cycles relative to an arbitrary origin.\label{def:time}
\item A clock cycles with time $t$ at a temporal frequency $f_c$.\label{def:clock}
\item Temporal frequency $f$ is measured by counting cycles per interval in time $t$.\label{def:temporalfrequency}
\item Position $\mathbf{r}$ is measured by counting ruler cycles relative to an arbitrary origin. \label{def:position}
\item A ruler cycles with position $\mathbf{r}$ at the spatial frequency $\mathbf{k}_c$ of a massless ($m=0$) wave cycling in time $t$ at the clock frequency $f_c$.\label{def:ruler}
\item Spatial frequency $\mathbf{k}$ is measured by counting cycles per interval in position $\mathbf{r}$.\label{def:spatialfrequency}
\item  The interval $s$ separating two events, one at time $t$ and position $\mathbf{r}$ and the other defined to be the origin, is given by $s^2=t^2-\mathbf{r}^2$.\label{def:interval}
\item At time $t$ and position $\mathbf{r}$, the phase $\phi$ of a wave with constant temporal frequency $f$ and spatial frequency $\mathbf{k}$ is given by $\phi=\mathbf{k}\cdot\mathbf{r}-ft$.\label{def:phase}
\item For a wave with temporal frequency $f$ and spatial frequency $\mathbf{k}$, the mass $m$ is given by $m^2=f^2-\mathbf{k}^2$.\label{def:mass}
\item A wave is an object that can be equivalently described as a function of time $t$ and position $\mathbf{r}$ or as a function of temporal frequency $f$ and spatial frequency $\mathbf{k}$.\label{def:wave}
\item A particle is an object that can be counted as an integer unit.  \label{def:particle}
\end{enumerate}
To keep these definitions as self-contained as possible, we use ahistorical units where $h=c=1$ (which avoids having to define `light') and avoid jargony shorthand (``A wave is an object with a Fourier transform.''). The definitions are necessarily self-referential (i.e. circular), but not completely tautological (``A wave is an object that satisfies a wave equation.''). They are also incomplete, but the undefined terms (e.g. `function') are logical or mathematical. Measuring is defined implicitly to be counting.\footnote{The frequency definitions \#\ref{def:temporalfrequency} and  \#\ref{def:spatialfrequency} recall a point raised earlier: classical physics combines the commonplace with the inaccessible.  These definitions are common and uncontroversial (which is to say mathematical) enough that most physics textbooks do not bother to discuss them. In a classical context they are both theoretically essential (from the standpoint presented here) and, often, experimentally impracticable.  See  Section~\ref{sec:scales} and the discussion of footballs in Section~\ref{sec:waves}.} Objects exhibit wave/particle duality: they can have wave properties and particle properties \cite{2000Harrison,1968Saxon,1999Espagnat}.

Definitions \#\ref{def:time}--\ref{def:particle} define the real-space arena (Section~\ref{sec:spacetime}), the reciprocal-space arena dual to the real-space arena, and the arena content. Putting waves (definition \#\ref{def:wave}) in the arena via the wave postulate (Section~\ref{sec:waves}) generates all of the `classical' dynamical laws (Sections~\ref{sec:waves}--\ref{sec:stationary}).  

These definitions fail to explain two asymmetries that are obvious to our unaided senses. First, we experience the real-space arena, not the reciprocal-space arena.  Perhaps this asymmetry is `by definition', but these definitions do not explain its origin.  Second, we experience the real-space arena asymmetrically: forward and backward directions in space are similar, while forward and backward directions in time are not. In fact, one can argue that time's \emph{defining} characteristic is that it orders events \cite{2018tHooft}. While these definitions \#\ref{def:time}--\ref{def:particle} define time separately from space, and include distinguishing minus signs in the invariants, they and the laws of physics that follow (Sections~\ref{sec:spacetime}--\ref{sec:stationary}) do not explain why one spacetime coordinate so distinguishes before, now, and after (or past, present, and future), and the other three do not. The so-called arrow of time is not understood at any level \cite{1997Savitt,1997Schulman}. 

Here the defining characteristic of a particle is countability, as opposed to localizability \cite{1999Espagnat}.  In a double-slit experiment, the diffracting object manifests its particle nature when it hits the screen because at that instant it can be counted, not because it is localized to a point.   Attempting to count the diffracting objects beforehand, for instance at the slit, destroys their (complementary) wave properties and the diffraction pattern \cite{1968Saxon}. Capturing the particle on the screen does constitute a measurement of the particle's position, but such measurements are better considered as counting the particle's presence or absence relative to a pre-measured coordinate system than as counting ruler cycles from the origin to where the particle is counted. When we capture a diffracting photon with a digital camera sensor, for instance, the position measurement answers the question, ``which pixel registered the count?'', not ``where is the pixel that registered the count?'', which presumably is known from the manufacturer's datasheet.

Localizing is, of course, a great aid to counting.  The strong association of these concepts leads to their conflation.  However, the definition given here is preferred because it can accommodate, for instance, a description of a particle in a delocalized state, which the usual definition cannot.  Counting is thus associated with normalizability.   In the double slit experiment, a particle can be launched at the slits and can hit the screen, for the initial and final states can be integrated to an integer value (unity).  However, a particle does not go through one slit or the other, as it cannot be counted (i.e. normalized) at either one. 

The blackbody problem (Appendix~\ref{sec:quantum}) gives another example --- one of great historical importance \cite{1966Jammer} --- that illustrates how particles are more properly defined as countable, as opposed to localizable. Considering thermal radiation as classical electromagnetic waves, which have no normalization constraint on their amplitudes, leads to the ultraviolet catastrophe. Assigning a countable nature to these waves, such that  electromagnetic modes are occupied in integer units, gives Planck's spectrum and agreement with experiment.  It is not required that the thermal radiation be localized.  In fact, often the thermal radiation in a blackbody cavity is analyzed in terms of standing waves, which are  delocalized.  Thus the blackbody problem is solved by postulating that the thermal electromagnetic field consists of countable particles (i.e. photons). 

The particle definition (\#\ref{def:particle}) leads to one more perspective on the classical limit and allows us to clarify the footnote in Table~\ref{table:dictionary}. We imagine a particle described by a Gaussian wavepacket.  In one dimension the wavepacket is specified by three quantities: its mean position in real space $x'$, its mean position in reciprocal space $k'$, and its width $\sigma_k$ in reciprocal space,
\begin{equation}\label{eq:Gaussian_uncertainty}
\sigma_k=\frac{1}{4 \pi \sigma_x},
\end{equation}
where $\sigma_x$ is the wavepacket's corresponding width in real space. Normalizing the wavepacket establishes the correspondence with the integers required to make the particle `countable'.  

This wavepacket can be constructed by shifting and then boosting a wavepacket initially centered on the real- and reciprocal-space origins. It can be described equally well in real space: 
\begin{equation}\label{psireal}
\psi^g(x;\sigma_k,x',k') = e^{2\pi i  k' x }\sqrt{2 \sigma_k \sqrt{2 \pi }} e^{-(2 \pi \sigma_k (x-x'))^2},
\end{equation}
or reciprocal space:
\begin{equation}\label{psireciprocal}
 \bar{\psi}^g(k;\sigma_k,x',k') = e^{-2 \pi  i (k-k') x'} \frac{1}{\sqrt{ \sigma_k\sqrt{2 \pi}}} e^{-(\frac{k-k'}{2\sigma_k})^2} 
.
\end{equation}

Now we wish to combine two particles to make a minimal, composite object.  Normally we imagine assembling objects additively.  For instance, we might add beans to a bag to make a beanbag, or we might add an oxygen atom to a carbon atom to make a carbon monoxide molecule.  According to the wave picture, composite objects are assembled not by addition, but by multiplication.  Just as amplitudes for events occurring in succession multiply \cite{2010Feynman}, so amplitudes for particles comprising an object multiply. In other words, we do not add the beans and the bag, we multiply the beans and the bag. 

That multiplication, not addition, describes the combination of two particles is a consequence of definition~\#\ref{def:particle}, which is the definition that introduces the `quantum'.  Classical waves, such as those on a guitar string, a drum-head, or a wind-swept sea, add with no particular restrictions on their amplitudes.  Definitions~\#\ref{def:time}--\ref{def:wave} can accommodate such waves, which show no tendency to become classical objects as the number of waves $N$ increases. In fact, without definition~\#\ref{def:particle} there is no number $N$. 

Quantum mechanical superposition is fundamentally different than classical superposition \cite{1967Dirac}, and this difference is essential for realizing the classical limit. Definition~\#\ref{def:particle} requires that each particle have its own space in the ledger, so that it is normalizable, which is to say that it can be counted.\footnote{We are making the usual classical assumption that the particles are distinguishable \cite{1968Schiff}.  If the particles are indistinguishable then `different' particles can superpose and interfere \cite{1974HanburyBrown,1956Purcell}.}   Superposition in the sense of addition occurs with the various states of one particle, but not with states of different particles. The `quantum' countability definition \#\ref{def:particle} creates $N$, and, as we will now show, the large-$N$ limit gives the wave-objects that we recognize as `classical' particles.\footnote{Feynman argues \cite{1964Feynman} that the sentence ``all things are made of atoms'' condenses much of our scientific knowledge into very few words. ``All things are made of countable waves'' is written in the same spirit, and, for the price of one additional word, communicates a good deal more.}

Initially we consider the two particles to be non-interacting, in which case their Hilbert spaces are not coupled. Combining the spaces in a tensor product corresponds to simple multiplication in both real- and reciprocal-space, so a minimal object consisting of two particles ($\psi^g_1(x_1;\sigma_1,x'_1,k'_1)$ and $\psi^g_1(x_2;\sigma_2,x'_2,k'_2)$) can be written $\psi^g_1(x_1;\sigma_1,x'_1,k'_1)\cdot\psi^g_1(x_2;\sigma_2,x'_2,k'_2)$ in a real-space representation.   Some algebra shows that this composite state has a second factorization (one of an infinite number),
\begin{align}\label{ObjectFromTwoParticles}
\psi^g(x_1;\sigma_1,x'_1,k'_1)&\cdot \psi^g(x_2;\sigma_2,x'_2,k'_2)= \\ \nonumber
& \qquad \psi^g(x;\sigma_k,x',k')\cdot  \psi^g(X;\sigma_K,X',K'),
\end{align}
where the coordinate transformation $(x_1,x_2)\rightarrow (x,X)$ is given by
\begin{equation}\label{transformations}
x=\frac{\sigma _1^2 x_1+\sigma _2^2 x_2}{\sigma _1^2+\sigma _2^2},\,\text{and }\quad X = x_1-x_2,
\end{equation}
and the Gaussian parameters by
\begin{align}\label{GaussianParameters}
x'&=\frac{\sigma_1^2 x_1'+\sigma_2^2 x_2'}{\sigma_1^2+\sigma_2^2},& X' &=x'_1-x'_2,\\ \nonumber
\sigma_k& =\sqrt{\sigma_1^2+\sigma_2^2},&\sigma_K&=\frac{\sigma_1 \sigma_2}{\sqrt{\sigma_1^2+\sigma_2^2}},\\ \nonumber
k'&=k'_1+k'_2,\quad \text{and}& K' &= \frac{\sigma_2^2 k_1'-\sigma_1^2 k_2'}{\sigma_1^2+\sigma_2^2}.
\end{align}
The formulas for the inverse transformation are
\begin{align}
x_{1,2}&=x\pm\frac{X }{2}  \left(1\mp\sqrt{1-\frac{4 \sigma_K^2}{\sigma_k ^2}}\right),\\
x'_{1,2}&=x'\pm \frac{X'}{2}  \left(1\mp\sqrt{1-\frac{4 \sigma_K^2}{\sigma_k ^2}}\right),\\
\sigma_{1,2}&=\sigma_k  \sqrt{\frac{1}{2} \left(1\pm\sqrt{1-\frac{4 \sigma_K^2}{\sigma_k ^2}}\right)},\,\text{and}\\
k'_{1,2}&=\frac{k'}{2}  \left(1\pm\sqrt{1-\frac{4 \sigma_K^2}{\sigma_k ^2}}\right)\pm K',
\end{align}
where the upper (lower) sign is chosen for particle 1 (2). If the particles do not have identical widths the axes scale, so the coordinate transformations \ref{transformations} are generally more than a simple rotation, but such stretching and compressing is permissible in this abstract vector space (as opposed to actual physical space) \cite{1971Symon}.  

Using an obvious shorthand, if $\psi^g(x_1)$ and $\psi^g(x_2)$ describe the first and second particles, respectively, then $\psi^g(x)$ and $\psi^g(X)$ can be taken to describe `external' and `internal' degrees-of-freedom, respectively (i.e. `extracules' and `intracules' \cite{1953Eddington,1967Coleman,2003Gill,2010Proud}). We also note that, if the $\sigma$'s are proportional to the square-roots of their respective masses, then the `external' and `internal' degrees-of-freedom may be associated with the center of mass and the reduced mass, respectively. At this point this connection is unexpected --- the Gaussian widths seem to be  parameters that are adjustable independent of the dispersion relation, and the dimensions of $\sigma_k^2$ and $m$ differ by a length scale.  We will return to examine this connection shortly.

In a standard quantum mechanical treatment, the internal degrees of freedom are of interest and the external coordinates are ignored.  Here, in order to understand how classical objects arise from waves, we take the opposite approach.  We integrate over all values of the internal coordinate $X$. The (normalized) wavefunction $\psi^g(X)$ then drops out of the problem. The object remaining now has a (Gaussian) particle description in terms of the `external' wavefunction $\psi^g(x)$ centered in reciprocal space on the sum wavevector $k'$ (i.e. momentum) and in real space between the constituent particles at $x'$. 

A complex, composite object can be built up by adding one particle after another, which corresponds to repeating this multiplication process again and again. By induction \cite{2018Bromiley} we find that the Gaussian parameters describing the final, external degree of freedom are
\begin{align}\label{GaussianObject}
\sigma_k^2& =\sum_{i=1}^N\sigma_{ki}^2,\\ \nonumber
x'&=\frac{1}{\sigma_k^2}\sum_{i=1}^N \sigma_{ki}^2 x'_i, \quad \text{and}\\ \nonumber
k'&=\sum_{i=1}^N k'_i.
\end{align}
The assembly process is similar in reciprocal-space and gives the same result as that obtained by Fourier transforming the final expression in its real-space representation.  

This path already looks promising, in that it reproduces  properties expected for an object made of multiple particles. For instance, in real space the composite object is centered around a weighted mean of the centers of its constituents. And in reciprocal space the wavevector (i.e. momentum) of the composite object is equal to the sum of the wavevectors of its constituents.

We now consider the case where the $N$ constituent particles are all similar.  For instance, the beans of the composite beanbag are moving together on a ballistic trajectory. (Once the beanbag is launched, each bean's path is approximately unaltered by the presence of its neighbors, which illustrates how individual particles with negligible interactions can still be grouped as a composite object.)  Then
 $\sigma_k \simeq \sqrt{N}\sigma_{ki}$, $k' \simeq N k'_i$, and $x'$ is the mean of the $x'_i$. Calculating the probability density for this composite object, we see that 
\begin{align}\label{realdensity}
\psi^{g*}(x)\cdot \psi^{g*}(x) &= 2 \sigma_k \sqrt{2 \pi } e^{-2(2 \pi  \sigma_k (x-x'))^2}\\
&\rightarrow \delta(x-x')\nonumber
\end{align}
in the limit that the number of constituent particles $N\rightarrow \infty$ because $\sigma_k\rightarrow \infty$. Thus the real-space probability density for a many-particle object approaches a $\delta$-function with a real-space width $\sigma_x \propto 1/\sqrt{N}$.  

In reciprocal space the argument is again more subtle.  The function
\begin{equation} \label{recipdensity}
 \bar{\psi}^{g*}(k)\cdot \bar{\psi}^{g}(k) = \frac{1}{ \sigma_k\sqrt{2 \pi}} e^{-2(\frac{k-k'}{2\sigma_k})^2} 
\end{equation}
does not approach a delta function because $\sigma_k \propto \sqrt{N}$ is becoming large.  However, $k' \propto N$ is becoming large faster, so that the function \ref{recipdensity} looks like a delta function from a perspective that encompasses both the reciprocal space origin and the peak.  From this perspective, the natural perspective for the problem, the relative width $\sigma_k/k'$ is decreasing $\propto 1/\sqrt{N}$ in reciprocal space, just as it is in real space.  Thus the reference to the `zoom level' in Section~\ref{sec:waves} is made precise: a wave can be well-localized in both real- and reciprocal-space without violating the uncertainty principle. 

According to definition \#\ref{def:wave}, a description of the wave in either real- or reciprocal-space can be converted into an equivalent description in the conjugate space.  In the classical limit, the phase becomes uncountably large and this connection is lost from a practical (i.e. experimental) standpoint.  In our example, either Eq.~\ref{psireal} or Eq.~\ref{psireciprocal} provides a complete description in terms of $x'$, $k'$, and a width (either $\sigma_x$ or $\sigma_k$). In the classical limit, the phase $\sim k'x'\propto N$ is unmeasurable\footnote{It is more accurate to say that the phase discussed here is unmeasurable not because it is large, but because it changes rapidly when varied separately in either time or space. The phase discussed here is only a (spatial) piece of the `full' phase --- Eq.~\ref{eq:phi} or Eq.~\ref{eq:goodaction} --- which is unchanging along the classical trajectory. (By choice of origin the phase might be set to zero, in which case it is not large at all.) Thus as we move into the classical limit the `full' phase controls the observable dynamics --- Section~\ref{sec:stationary} --- while the `partial' phases, i.e. the leading terms in Eqs.~\ref{psireal}--\ref{psireciprocal}, are becoming unobservable.  Also note that, while unobservable, these phases are not incoherent.  The phase relationships, i.e. coherences, are what maintain the object as a unified whole.  Fiddling with a particle's phase changes its position or its wavevector/momentum (Eqs.~\ref{psireal}--\ref{psireciprocal}).  To the extent that these physical quantities are constrained in a composite object, the phases must be coherent. Coherent states of the `quantum' harmonic oscillator (Appendix~\ref{sec:quantum}) provide a concrete example of the coherence of `classical' objects.}, leaving Eqs.~\ref{realdensity}--\ref{recipdensity} as the information available. These equations are effectively $\delta(x-x')$ and $\delta(k-k')$ respectively, and knowledge of one provides no information about the other.  In the classical limit, the position and the spatial frequency (i.e. momentum) descriptions are effectively decoupled.

Here we can also get a more quantitative sense of how the inequalities \ref{inequalities} develop as we move into the classical limit.  For a 1D, $N$-particle composite object, we can re-write
\begin{equation*}\tag{\ref{inequalities}}
\MyFourTerm
\end{equation*}
as
\begin{equation}
N \sigma_{xi} \gg \sigma_x \gg \frac{1}{4 \pi k'} \gg \frac{\sigma_k}{4 \pi k'^2},
\end{equation}
where the replacements are justified by arguments given previously, except for $x\rightarrow N \sigma_{xi}$.  This replacement bounds the size of the object by requiring that its constituent particles occupy distinct regions of space\footnote{While the spatial uncertainty relation (Eq.~\ref{goodhup}) has a direct analog $\Delta t \Delta f \geq 1/4\pi$, and $k\gg \Delta k$ implies $f\gg \Delta f$,  this replacement and $x\gg \Delta x$ do not have such obvious temporal analogs. Objects do not overlap in space, but they do in time. Despite the apparent similarity of Eqs.~\ref{eq:timeFT} and Eqs.~\ref{eq:spaceFT}, the uncertainties in the variables $t$ and $x$ seem to require very different interpretations.  Usually we assume that a measurement of $x$ returning one particular value in the range $\Delta x$ thereby excludes others, while we expect that a particle occupies all $t$ in $\Delta t$ \cite{1985Mermin}.}.  Justifying this bound requires ideas  (e.g. Coulomb repulsion and the Pauli exclusion principle) not discussed here, but that are probably familiar.  Using the uncertainty relation for Gaussians (Eq.~\ref{eq:Gaussian_uncertainty}) and the relations that follow from \ref{GaussianObject} gives
\begin{equation}
\frac{N}{4\pi\sigma_\text{ki} }\gg \frac{1}{4\pi\sqrt{N}\sigma_\text{ki}} \gg \frac{1}{4 \pi N k_i'} \gg \frac{\sigma_\text{ki}}{4 \pi \sqrt{N^3}k_i'^2}.
\end{equation}
Multiplying through by $4\pi\sigma_\text{ki}$ puts this inequality chain in a dimensionless form,  
\begin{equation}
N\gg \frac{1}{\sqrt{N}} \gg  \big(\frac{\sigma_\text{ki}}{k_i'}\big)\frac{1}{N} \gg \big(\frac{\sigma_\text{ki}}{k_i'}\big)^2\frac{1}{N\sqrt{N}}.
\end{equation}
These four terms scale like $N^1$, $N^{-1/2}$, $N^{-1}$, and $N^{-3/2}$, respectively. Thus as the number of particles $N \rightarrow \infty$, the classical limit is inevitable, provided that  $\sigma_\text{ki}\gg k'_i$ is not the case.  If $k'_i$ is not obviously appreciable, the thermal arguments given earlier in Appendix~\ref{sec:quantum} and here in this Appendix~\ref{particles} can be applied to determine whether the classical limit obtains.

In assembling the composite object we have neglected interactions between the constituent particles.  However, reviewing this development we see that, because we integrate over the internal degrees of freedom, the exact forms of the internal wavefunctions are unimportant. For instance, if the minimal, two-particle object is a hydrogen atom consisting of a proton and an electron, the Coulomb interaction gives internal coordinates described by hydrogenic wavefunctions.  These wavefunctions are not of a Gaussian form $\psi^g$ (Eqs.~\ref{psireal}--\ref{psireciprocal}).  In this case the states can only be factored in the external/internal coordinate system ($\mathbf{r},\,\mathbf{R}$), for in the electron/proton coordinate system  ($\mathbf{r}_\text{e},\,\mathbf{r}_\text{p}$) the states are entangled.  However, our goal here is see how a single object arises by combining many particles, not to specify how the particles arise by decomposing the object. The composite object's state need not be factorizable in terms of the constituent particle coordinates. We only require the factorization on right-hand side of Eq.~\ref{ObjectFromTwoParticles}. Often the transformation necessary to produce this separation of variables can be found, with the transformation to  center of mass and reduced mass coordinates being the classic example \cite{2010Feynman,1986Ghirardi}. In fact, in the many particle limit it is not even necessary that the initial \emph{external} wavefunctions be Gaussian. The central limit theorem \cite{2000Bracewell} ensures that, with suitably well-behaved constituents\footnote{A double slit or a grating, for instance, can induce poor behavior.}, the external wavefunction will eventually adopt a Gaussian form as the number of constituent particles increases, regardless of the specifics of the problem.

Thus this argument showing how classical objects (localized in real and reciprocal space) arise from waves (localized in neither space) is more general than initially advertised.  The passage to the classical limit is not restricted to non-interacting particles described by Gaussian wavepackets, but is, as it must be, a process that generally occurs with large numbers of particles, regardless of their interactions.

To recapitulate while generalizing to three dimensions: we start with $N$ wavepackets, each of which is described in its own real and reciprocal spaces in terms of its  mean position $\mathbf{r'}_i$, mean wavevector $\mathbf{k'}_i$, and reciprocal-space widths $\mytensor{\boldsymbol{\sigma}}_{ki}$.  These $N$ wavepackets represent particles that might be the constituent beans of a beanbag, the atoms that make up the beans, or the fundamental particles that make up the atoms. We rearrange the coordinates of the $3N$-dimensional space and integrate over the $3(N-1)$ `internal' coordinates. The `internal' coordinates then drop out of the problem, one by one, because the normalization conditions from each of the constituent particles have been transferred to the new degrees of freedom (i.e. pseudo-particles).  The remaining `external' 3D space contains a single normalized wavepacket representing the composite (pseudo-)particle. In real space this wavepacket is centered \emph{between} the constituent wavepackets' positions, and  in reciprocal space it is centered \emph{on the sum} of the constituent wavepacket's positions. For large $N$ this final wavepacket's widths are negligible in both real- and reciprocal space, where the bases for comparison are the object's size and its total wavevector, respectively. The Fourier connection between the real- and reciprocal-space descriptions is practically severed, and the resulting object can be treated like a `particle' in the sense of the word as employed by standard textbooks on classical mechanics. 

Returning to review the arc of the argument from its beginning (Section~\ref{sec:waves}), eliminating the `uninteresting' degrees of freedom from a complex, large-$N$ collection of particles produces a single `particle' that is well-localized (though not small) and behaves classically. In real space its location is effectively described by a delta-function $\sim \delta(\mathbf{r}-\mathbf{r'})$, where $\mathbf{r'}$ develops in time according to Eq.~\ref{eq:goodHamilton_groupv}.  In reciprocal space its location is effectively described by a delta-function $\sim \delta(\mathbf{k}-\mathbf{k'})$, where $\mathbf{k'}$ develops in time according to Eq.~\ref{eq:goodHamilton2ndLaw}. Thus the complete motion of the wavepacket is dictated by the condition that the phase $\phi=\mathbf{k}\cdot\mathbf{r}-f t$ is stationary.  In traditional language, one says that the motion of the object is dictated by the condition that the action $S$ is stationary. This `stationary action' condition is abstract and unmotivated, while the wave formulation's  `stationary phase' is comparatively elementary and might even supply a mental picture --- see, for example, the case of a massless object (Eq.~\ref{eq:genericwave}), where `stationary' becomes `constant'.

The relationship between the average position of a Gaussian wavepacket in real-space and the center of mass remains obscure. Proportionality between $\sigma_k^2$ and mass $m$ is seen in some important and basic contexts. It appears in the propagator for a free particle, where $\sigma_k^2=-im/(4\pi\Delta t)$ and $\Delta t$ is the time interval of propagation \cite{2010Feynman}. And it appears in the ground state of a harmonic oscillator, where $\sigma_{K}^2= \mu f_0/2$, $\mu$ is the reduced mass, and $f_0$ is the oscillator's frequency.  In other words, in the position representation (with $h=c=1$) a 1D harmonic oscillator's ground state $\psi^\text{ho}$ can be written
\begin{align}
\psi^\text{ho}(X) &= (4\pi \mu f_0)^{1/4}e^{-2 \pi^2 \mu f_0 X^2}\\ \nonumber
&=\psi^g(X;\sigma_{K},0,0).
\end{align}
This Gaussian state gives $\sigma_{K}^2$ proportional to the correct mass.  It also has a parameter ($f_0$) that can be adjusted to fit the problem: increasing the oscillator frequency localizes the state in real-space and delocalizes it in reciprocal-space, and vice-versa. However, for $x'$ to be equivalent to the center of mass, we require that $m_1$, $m_2$, the total mass $m=m_1+m_2$, and the reduced mass $\mu=m_1 m_2/(m_1+m_2)$ all be proportional to their respective $\sigma_k^2$'s with the \emph{same} constant of proportionality (\ref{GaussianParameters}). This requirement seems to over-constrain the problem.  For example, a carbon monoxide molecule in its vibrational ground state might be trapped in a harmonic potential, with no specific relationship between the  molecular binding and the trapping potential's depth.

The center of mass concept itself is not without problems.  It is most useful in situations with a great deal of symmetry (e.g. spherical), where the equivalence with the Gaussian center will probably hold. Without such symmetry a center of mass decomposition with as few as three particles is only approximate, and distinctions arise between the center of mass and the center of gravity \cite{1971Symon}. Moreover, the center of mass concept is seriously limited in that it does not have an obvious relativistic generalization \cite{1948Pryce,1949Newton,2007Alba}.  Thus the lack of an exact agreement between  $x'$ and the center of mass is intriguing, but not so disturbing that it shakes our confidence in the general soundness of these arguments.

\bibliography{hbar}

\begin{thebibliography}{102}%
\makeatletter
\providecommand \@ifxundefined [1]{%
 \@ifx{#1\undefined}
}%
\providecommand \@ifnum [1]{%
 \ifnum #1\expandafter \@firstoftwo
 \else \expandafter \@secondoftwo
 \fi
}%
\providecommand \@ifx [1]{%
 \ifx #1\expandafter \@firstoftwo
 \else \expandafter \@secondoftwo
 \fi
}%
\providecommand \natexlab [1]{#1}%
\providecommand \enquote  [1]{``#1''}%
\providecommand \bibnamefont  [1]{#1}%
\providecommand \bibfnamefont [1]{#1}%
\providecommand \citenamefont [1]{#1}%
\providecommand \href@noop [0]{\@secondoftwo}%
\providecommand \href [0]{\begingroup \@sanitize@url \@href}%
\providecommand \@href[1]{\@@startlink{#1}\@@href}%
\providecommand \@@href[1]{\endgroup#1\@@endlink}%
\providecommand \@sanitize@url [0]{\catcode `\\12\catcode `\$12\catcode
  `\&12\catcode `\#12\catcode `\^12\catcode `\_12\catcode `\%12\relax}%
\providecommand \@@startlink[1]{}%
\providecommand \@@endlink[0]{}%
\providecommand \url  [0]{\begingroup\@sanitize@url \@url }%
\providecommand \@url [1]{\endgroup\@href {#1}{\urlprefix }}%
\providecommand \urlprefix  [0]{URL }%
\providecommand \Eprint [0]{\href }%
\providecommand \doibase [0]{http://dx.doi.org/}%
\providecommand \selectlanguage [0]{\@gobble}%
\providecommand \bibinfo  [0]{\@secondoftwo}%
\providecommand \bibfield  [0]{\@secondoftwo}%
\providecommand \translation [1]{[#1]}%
\providecommand \BibitemOpen [0]{}%
\providecommand \bibitemStop [0]{}%
\providecommand \bibitemNoStop [0]{.\EOS\space}%
\providecommand \EOS [0]{\spacefactor3000\relax}%
\providecommand \BibitemShut  [1]{\csname bibitem#1\endcsname}%
\let\auto@bib@innerbib\@empty
\bibitem [{\citenamefont {Abers}(2004)}]{2004Abers}%
  \BibitemOpen
  \bibfield  {author} {\bibinfo {author} {\bibnamefont {Abers}, \bibfnamefont
  {Ernest~S}}} (\bibinfo {year} {2004}),\ \href@noop {} {\emph {\bibinfo
  {title} {Quantum Mechanics}}}\ (\bibinfo  {publisher} {{Pearson Education}},\
  \bibinfo {address} {{Upper Saddle River, N.J.}})\BibitemShut {NoStop}%
\bibitem [{\citenamefont {Alba}\ \emph {et~al.}(2007)\citenamefont {Alba},
  \citenamefont {Crater},\ and\ \citenamefont {Lusanna}}]{2007Alba}%
  \BibitemOpen
  \bibfield  {author} {\bibinfo {author} {\bibnamefont {Alba}, \bibfnamefont
  {David}}, \bibinfo {author} {\bibfnamefont {Horace~W.}\ \bibnamefont
  {Crater}}, \ and\ \bibinfo {author} {\bibfnamefont {Luca}\ \bibnamefont
  {Lusanna}}} (\bibinfo {year} {2007}),\ \bibfield  {title} {\enquote {\bibinfo
  {title} {Hamiltonian relativistic two-body problem: Center of mass and orbit
  reconstruction},}\ }\href@noop {} {\bibfield  {journal} {\bibinfo  {journal}
  {Journal of Physics A: Mathematical and Theoretical}\ }\textbf {\bibinfo
  {volume} {40}}~(\bibinfo {number} {31}),\ \bibinfo {pages}
  {9585--9607}}\BibitemShut {NoStop}%
\bibitem [{\citenamefont {Bassi}\ and\ \citenamefont
  {Ghirardi}(2003)}]{2003Bassi}%
  \BibitemOpen
  \bibfield  {author} {\bibinfo {author} {\bibnamefont {Bassi}, \bibfnamefont
  {Angelo}}, \ and\ \bibinfo {author} {\bibfnamefont {GianCarlo}\ \bibnamefont
  {Ghirardi}}} (\bibinfo {year} {2003}),\ \bibfield  {title} {\enquote
  {\bibinfo {title} {Dynamical reduction models},}\ }\href@noop {} {\bibfield
  {journal} {\bibinfo  {journal} {Physics Reports}\ }\textbf {\bibinfo {volume}
  {379}}~(\bibinfo {number} {5}),\ \bibinfo {pages} {257--426}}\BibitemShut
  {NoStop}%
\bibitem [{\citenamefont {Bord{\'e}}(2005)}]{2005Borde}%
  \BibitemOpen
  \bibfield  {author} {\bibinfo {author} {\bibnamefont {Bord{\'e}},
  \bibfnamefont {Christian~J}}} (\bibinfo {year} {2005}),\ \bibfield  {title}
  {\enquote {\bibinfo {title} {Base units of the {{SI}}, fundamental constants
  and modern quantum physics},}\ }\href@noop {} {\bibfield  {journal} {\bibinfo
   {journal} {Philosophical Transactions of the Royal Society A: Mathematical,
  Physical and Engineering Sciences}\ }\textbf {\bibinfo {volume}
  {363}}~(\bibinfo {number} {1834}),\ \bibinfo {pages}
  {2177--2201}}\BibitemShut {NoStop}%
\bibitem [{\citenamefont {Bracewell}(2000)}]{2000Bracewell}%
  \BibitemOpen
  \bibfield  {author} {\bibinfo {author} {\bibnamefont {Bracewell},
  \bibfnamefont {Ronald~N}}} (\bibinfo {year} {2000}),\ \href@noop {} {\emph
  {\bibinfo {title} {The {{Fourier}} Transform and Its Applications}}},\
  \bibinfo {edition} {3rd}\ ed.,\ {{McGraw}}-{{Hill}} Series in Electrical and
  Computer Engineering. {{Circuits}} and Systems\ (\bibinfo  {publisher}
  {{McGraw Hill}},\ \bibinfo {address} {{Boston}})\BibitemShut {NoStop}%
\bibitem [{\citenamefont {Bridgman}(1931)}]{1931Bridgman}%
  \BibitemOpen
  \bibfield  {author} {\bibinfo {author} {\bibnamefont {Bridgman},
  \bibfnamefont {P~W}}} (\bibinfo {year} {1931}),\ \href@noop {} {\emph
  {\bibinfo {title} {Dimensional Analysis}}}\ (\bibinfo  {publisher} {{Yale
  University Press}},\ \bibinfo {address} {{New Haven}})\BibitemShut {NoStop}%
\bibitem [{\citenamefont {Bromiley}(2018)}]{2018Bromiley}%
  \BibitemOpen
  \bibfield  {author} {\bibinfo {author} {\bibnamefont {Bromiley},
  \bibfnamefont {Paul~A}}} (\bibinfo {year} {2018}),\ \href@noop {} {\emph
  {\bibinfo {title} {Products and {{Convolutions}} of {{Gaussian Probability
  Density Functions}}}}},\ \bibinfo {type} {Internal Report}\ \bibinfo {number}
  {2003-003}\ (\bibinfo  {institution} {{School of Health Sciences, University
  of Manchester}},\ \bibinfo {address} {{Division of Informatics, Imaging and
  Data Sciences}})\BibitemShut {NoStop}%
\bibitem [{\citenamefont {Cheng}\ and\ \citenamefont {Li}(1991)}]{1991Cheng}%
  \BibitemOpen
  \bibfield  {author} {\bibinfo {author} {\bibnamefont {Cheng}, \bibfnamefont
  {Ta-Pei}}, \ and\ \bibinfo {author} {\bibfnamefont {Ling-Fong.}\ \bibnamefont
  {Li}}} (\bibinfo {year} {1991}),\ \href@noop {} {\emph {\bibinfo {title}
  {Gauge Theory of Elementary Particle Physics}}},\ Oxford Science
  Publications\ (\bibinfo  {publisher} {{Clarendon Press}},\ \bibinfo {address}
  {{Oxford}})\BibitemShut {NoStop}%
\bibitem [{\citenamefont {Chiao}\ \emph {et~al.}(1989)\citenamefont {Chiao},
  \citenamefont {Hong}, \citenamefont {Kwiat}, \citenamefont {Nathel},\ and\
  \citenamefont {Vareka}}]{1989Chiao}%
  \BibitemOpen
  \bibfield  {author} {\bibinfo {author} {\bibnamefont {Chiao}, \bibfnamefont
  {R~Y}}, \bibinfo {author} {\bibfnamefont {C.~K.}\ \bibnamefont {Hong}},
  \bibinfo {author} {\bibfnamefont {P.~G.}\ \bibnamefont {Kwiat}}, \bibinfo
  {author} {\bibfnamefont {H.}~\bibnamefont {Nathel}}, \ and\ \bibinfo {author}
  {\bibfnamefont {W.~A.}\ \bibnamefont {Vareka}}} (\bibinfo {year} {1989}),\
  \bibfield  {title} {\enquote {\bibinfo {title} {Optical {{Manifestations}} of
  {{Berry}}'s {{Topological Phase}}: {{Classical}} and {{Quantum Aspects}}},}\
  }in\ \href@noop {} {\emph {\bibinfo {booktitle} {Coherence and {{Quantum
  Optics VI}}}}},\ \bibinfo {editor} {edited by\ \bibinfo {editor}
  {\bibfnamefont {Joseph~H.}\ \bibnamefont {Eberly}}, \bibinfo {editor}
  {\bibfnamefont {Leonard}\ \bibnamefont {Mandel}}, \ and\ \bibinfo {editor}
  {\bibfnamefont {Emil}\ \bibnamefont {Wolf}}}\ (\bibinfo  {publisher}
  {{Springer US}})\ pp.\ \bibinfo {pages} {155--160}\BibitemShut {NoStop}%
\bibitem [{\citenamefont {Coelho}(2012)}]{2012Coelho}%
  \BibitemOpen
  \bibfield  {author} {\bibinfo {author} {\bibnamefont {Coelho}, \bibfnamefont
  {Ricardo~Lopes}}} (\bibinfo {year} {2012}),\ \bibfield  {title} {\enquote
  {\bibinfo {title} {On the {{Definition}} of {{Mass}} in {{Mechanics}}: {{Why
  Is It So Difficult}}?}}\ }\href@noop {} {\bibfield  {journal} {\bibinfo
  {journal} {The Physics Teacher}\ }\textbf {\bibinfo {volume} {50}}~(\bibinfo
  {number} {5}),\ \bibinfo {pages} {304--306}}\BibitemShut {NoStop}%
\bibitem [{\citenamefont {Cohen}\ \emph {et~al.}(2019)\citenamefont {Cohen},
  \citenamefont {Larocque}, \citenamefont {Bouchard}, \citenamefont
  {Nejadsattari}, \citenamefont {Gefen},\ and\ \citenamefont
  {Karimi}}]{2019Cohen}%
  \BibitemOpen
  \bibfield  {author} {\bibinfo {author} {\bibnamefont {Cohen}, \bibfnamefont
  {Eliahu}}, \bibinfo {author} {\bibfnamefont {Hugo}\ \bibnamefont {Larocque}},
  \bibinfo {author} {\bibfnamefont {Fr{\'e}d{\'e}ric}\ \bibnamefont
  {Bouchard}}, \bibinfo {author} {\bibfnamefont {Farshad}\ \bibnamefont
  {Nejadsattari}}, \bibinfo {author} {\bibfnamefont {Yuval}\ \bibnamefont
  {Gefen}}, \ and\ \bibinfo {author} {\bibfnamefont {Ebrahim}\ \bibnamefont
  {Karimi}}} (\bibinfo {year} {2019}),\ \bibfield  {title} {\enquote {\bibinfo
  {title} {Geometric phase from {{Aharonov}}\textendash{{Bohm}} to
  {{Pancharatnam}}\textendash{{Berry}} and beyond},}\ }\href@noop {} {\bibfield
   {journal} {\bibinfo  {journal} {Nature Reviews Physics}\ }\textbf {\bibinfo
  {volume} {1}}~(\bibinfo {number} {7}),\ \bibinfo {pages}
  {437--449}}\BibitemShut {NoStop}%
\bibitem [{\citenamefont {{Cohen-Tannoudji}}\ \emph {et~al.}(1977)\citenamefont
  {{Cohen-Tannoudji}}, \citenamefont {Diu},\ and\ \citenamefont
  {Lalo{\"e}}}]{1977Cohen-Tannoudji}%
  \BibitemOpen
  \bibfield  {author} {\bibinfo {author} {\bibnamefont {{Cohen-Tannoudji}},
  \bibfnamefont {Claude}}, \bibinfo {author} {\bibfnamefont {Bernard}\
  \bibnamefont {Diu}}, \ and\ \bibinfo {author} {\bibfnamefont {Franck}\
  \bibnamefont {Lalo{\"e}}}} (\bibinfo {year} {1977}),\ \href@noop {} {\emph
  {\bibinfo {title} {Quantum Mechanics}}}\ (\bibinfo  {publisher} {{Wiley}},\
  \bibinfo {address} {{New York}})\BibitemShut {NoStop}%
\bibitem [{\citenamefont {Coleman}(1967)}]{1967Coleman}%
  \BibitemOpen
  \bibfield  {author} {\bibinfo {author} {\bibnamefont {Coleman}, \bibfnamefont
  {A~J}}} (\bibinfo {year} {1967}),\ \bibfield  {title} {\enquote {\bibinfo
  {title} {Density matrices in the quantum theory of matter: {{Energy}},
  intracules and extracules},}\ }\href@noop {} {\bibfield  {journal} {\bibinfo
  {journal} {International Journal of Quantum Chemistry}\ }\textbf {\bibinfo
  {volume} {1}}~(\bibinfo {number} {S1}),\ \bibinfo {pages}
  {457--464}}\BibitemShut {NoStop}%
\bibitem [{\citenamefont {Commins}(2014)}]{2014Commins}%
  \BibitemOpen
  \bibfield  {author} {\bibinfo {author} {\bibnamefont {Commins}, \bibfnamefont
  {Eugene~D}}} (\bibinfo {year} {2014}),\ \href@noop {} {\emph {\bibinfo
  {title} {Quantum Mechanics : An Experimentalist's Approach}}}\ (\bibinfo
  {publisher} {{Cambridge University Press}},\ \bibinfo {address} {{New York,
  NY}})\BibitemShut {NoStop}%
\bibitem [{\citenamefont {Cowley}(1995)}]{1995Cowley}%
  \BibitemOpen
  \bibfield  {author} {\bibinfo {author} {\bibnamefont {Cowley}, \bibfnamefont
  {J~M}}} (\bibinfo {year} {1995}),\ \href@noop {} {\emph {\bibinfo {title}
  {Diffraction Physics}}},\ \bibinfo {edition} {3rd}\ ed.,\ North-{{Holland}}
  Personal Library\ (\bibinfo  {publisher} {{Elsevier Science B.V.}},\ \bibinfo
  {address} {{Amsterdam}})\BibitemShut {NoStop}%
\bibitem [{\citenamefont {Crawford}(1968)}]{1968Crawford}%
  \BibitemOpen
  \bibfield  {author} {\bibinfo {author} {\bibnamefont {Crawford},
  \bibfnamefont {Frank~S}}} (\bibinfo {year} {1968}),\ \href@noop {} {\emph
  {\bibinfo {title} {Waves}}},\ Berkeley Physics Course, v. 3\ (\bibinfo
  {publisher} {{McGraw-Hill}},\ \bibinfo {address} {{New York}})\BibitemShut
  {NoStop}%
\bibitem [{\citenamefont {Cronin}\ \emph {et~al.}(2009)\citenamefont {Cronin},
  \citenamefont {Schmiedmayer},\ and\ \citenamefont {Pritchard}}]{2009Cronin}%
  \BibitemOpen
  \bibfield  {author} {\bibinfo {author} {\bibnamefont {Cronin}, \bibfnamefont
  {Alexander~D}}, \bibinfo {author} {\bibfnamefont {J{\"o}rg}\ \bibnamefont
  {Schmiedmayer}}, \ and\ \bibinfo {author} {\bibfnamefont {David~E.}\
  \bibnamefont {Pritchard}}} (\bibinfo {year} {2009}),\ \bibfield  {title}
  {\enquote {\bibinfo {title} {Optics and interferometry with atoms and
  molecules},}\ }\href@noop {} {\bibfield  {journal} {\bibinfo  {journal}
  {Reviews of Modern Physics}\ }\textbf {\bibinfo {volume} {81}}~(\bibinfo
  {number} {3}),\ \bibinfo {pages} {1051--1129}}\BibitemShut {NoStop}%
\bibitem [{\citenamefont {Dirac}(1933)}]{1933Dirac}%
  \BibitemOpen
  \bibfield  {author} {\bibinfo {author} {\bibnamefont {Dirac}, \bibfnamefont
  {P~A~M}}} (\bibinfo {year} {1933}),\ \bibfield  {title} {\enquote {\bibinfo
  {title} {The {{Lagrangian}} in {{Quantum Mechanics}}},}\ }\href@noop {}
  {\bibfield  {journal} {\bibinfo  {journal} {Physikalische Zeitschrift der
  Sowjetunion}\ }\textbf {\bibinfo {volume} {3}}~(\bibinfo {number} {1}),\
  \bibinfo {pages} {64--72}}\BibitemShut {NoStop}%
\bibitem [{\citenamefont {Dirac}(1967)}]{1967Dirac}%
  \BibitemOpen
  \bibfield  {author} {\bibinfo {author} {\bibnamefont {Dirac}, \bibfnamefont
  {P~A~M}}} (\bibinfo {year} {1967}),\ \href@noop {} {\emph {\bibinfo {title}
  {The Principles of Quantum Mechanics.}}},\ \bibinfo {edition} {4th}\ ed.\
  (\bibinfo  {publisher} {{Clarendon Press}},\ \bibinfo {address}
  {{Oxford}})\BibitemShut {NoStop}%
\bibitem [{\citenamefont {Duff}(2015)}]{2015Duff}%
  \BibitemOpen
  \bibfield  {author} {\bibinfo {author} {\bibnamefont {Duff}, \bibfnamefont
  {M~J}}} (\bibinfo {year} {2015}),\ \bibfield  {title} {\enquote {\bibinfo
  {title} {How fundamental are fundamental constants?}}\ }\href@noop {}
  {\bibfield  {journal} {\bibinfo  {journal} {Contemporary Physics}\ }\textbf
  {\bibinfo {volume} {56}}~(\bibinfo {number} {1}),\ \bibinfo {pages}
  {35--47}}\BibitemShut {NoStop}%
\bibitem [{\citenamefont {Eddington}(1928)}]{1928Eddington}%
  \BibitemOpen
  \bibfield  {author} {\bibinfo {author} {\bibnamefont {Eddington},
  \bibfnamefont {Arthur~S}}} (\bibinfo {year} {1928}),\ \href@noop {} {\emph
  {\bibinfo {title} {The {{Nature}} of the {{Physical World}} : The {{Gifford
  Lectures}} of 1927}}}\ (\bibinfo  {publisher} {{Cambridge University
  Press}})\BibitemShut {NoStop}%
\bibitem [{\citenamefont {Eddington}(1953)}]{1953Eddington}%
  \BibitemOpen
  \bibfield  {author} {\bibinfo {author} {\bibnamefont {Eddington},
  \bibfnamefont {Arthur Stanley~Sir}}} (\bibinfo {year} {1953}),\ \href@noop {}
  {\emph {\bibinfo {title} {Fundamental Theory}}},\ edited by\ \bibinfo
  {editor} {\bibfnamefont {E.~T.}\ \bibnamefont {Whittaker}}\ (\bibinfo
  {publisher} {{University Press}},\ \bibinfo {address} {{Cambridge
  [England]}})\BibitemShut {NoStop}%
\bibitem [{\citenamefont {Eisberg}\ and\ \citenamefont
  {Resnick}(1985)}]{1985Eisberg}%
  \BibitemOpen
  \bibfield  {author} {\bibinfo {author} {\bibnamefont {Eisberg}, \bibfnamefont
  {Robert~Martin}}, \ and\ \bibinfo {author} {\bibfnamefont {Robert}\
  \bibnamefont {Resnick}}} (\bibinfo {year} {1985}),\ \href@noop {} {\emph
  {\bibinfo {title} {Quantum Physics of Atoms, Molecules, Solids, Nuclei, and
  Particles}}},\ \bibinfo {edition} {2nd}\ ed.\ (\bibinfo  {publisher}
  {{Wiley}},\ \bibinfo {address} {{New York}})\BibitemShut {NoStop}%
\bibitem [{\citenamefont {d'~Espagnat}(1999)}]{1999Espagnat}%
  \BibitemOpen
  \bibfield  {author} {\bibinfo {author} {\bibnamefont {d'~Espagnat},
  \bibfnamefont {Bernard}}} (\bibinfo {year} {1999}),\ \href@noop {} {\emph
  {\bibinfo {title} {Conceptual Foundations of Quantum Mechanics}}},\ \bibinfo
  {edition} {2nd}\ ed.,\ Advanced Book Classics\ (\bibinfo  {publisher}
  {{Advanced Book Program, Perseus Books}},\ \bibinfo {address} {{Reading,
  Mass.}})\BibitemShut {NoStop}%
\bibitem [{\citenamefont {Ewald}(1969)}]{1969Ewald}%
  \BibitemOpen
  \bibfield  {author} {\bibinfo {author} {\bibnamefont {Ewald}, \bibfnamefont
  {P~P}}} (\bibinfo {year} {1969}),\ \bibfield  {title} {\enquote {\bibinfo
  {title} {Introduction to the dynamical theory of {{X}}-ray diffraction},}\
  }\href@noop {} {\bibfield  {journal} {\bibinfo  {journal} {Acta
  Crystallographica Section A: Crystal Physics, Diffraction, Theoretical and
  General Crystallography}\ }\textbf {\bibinfo {volume} {25}}~(\bibinfo
  {number} {1}),\ \bibinfo {pages} {103--108}}\BibitemShut {NoStop}%
\bibitem [{\citenamefont {Fein}\ \emph {et~al.}(2019)\citenamefont {Fein},
  \citenamefont {Geyer}, \citenamefont {Zwick}, \citenamefont {Kia{\l}ka},
  \citenamefont {Pedalino}, \citenamefont {Mayor}, \citenamefont {Gerlich},\
  and\ \citenamefont {Arndt}}]{2019Fein}%
  \BibitemOpen
  \bibfield  {author} {\bibinfo {author} {\bibnamefont {Fein}, \bibfnamefont
  {Yaakov~Y}}, \bibinfo {author} {\bibfnamefont {Philipp}\ \bibnamefont
  {Geyer}}, \bibinfo {author} {\bibfnamefont {Patrick}\ \bibnamefont {Zwick}},
  \bibinfo {author} {\bibfnamefont {Filip}\ \bibnamefont {Kia{\l}ka}}, \bibinfo
  {author} {\bibfnamefont {Sebastian}\ \bibnamefont {Pedalino}}, \bibinfo
  {author} {\bibfnamefont {Marcel}\ \bibnamefont {Mayor}}, \bibinfo {author}
  {\bibfnamefont {Stefan}\ \bibnamefont {Gerlich}}, \ and\ \bibinfo {author}
  {\bibfnamefont {Markus}\ \bibnamefont {Arndt}}} (\bibinfo {year} {2019}),\
  \bibfield  {title} {\enquote {\bibinfo {title} {Quantum superposition of
  molecules beyond 25 {{kDa}}},}\ }\href@noop {} {\bibfield  {journal}
  {\bibinfo  {journal} {Nature Physics}\ }\textbf {\bibinfo {volume}
  {15}}~(\bibinfo {number} {12}),\ \bibinfo {pages} {1242--1245}}\BibitemShut
  {NoStop}%
\bibitem [{\citenamefont {Feynman}(1948)}]{1948Feynman}%
  \BibitemOpen
  \bibfield  {author} {\bibinfo {author} {\bibnamefont {Feynman}, \bibfnamefont
  {R~P}}} (\bibinfo {year} {1948}),\ \bibfield  {title} {\enquote {\bibinfo
  {title} {Space-{{Time Approach}} to {{Non}}-{{Relativistic Quantum
  Mechanics}}},}\ }\href@noop {} {\bibfield  {journal} {\bibinfo  {journal}
  {Reviews of Modern Physics}\ }\textbf {\bibinfo {volume} {20}}~(\bibinfo
  {number} {2}),\ \bibinfo {pages} {367--387}}\BibitemShut {NoStop}%
\bibitem [{\citenamefont {Feynman}\ \emph {et~al.}(2010)\citenamefont
  {Feynman}, \citenamefont {Hibbs},\ and\ \citenamefont {Styer}}]{2010Feynman}%
  \BibitemOpen
  \bibfield  {author} {\bibinfo {author} {\bibnamefont {Feynman}, \bibfnamefont
  {Richard~P}}, \bibinfo {author} {\bibfnamefont {Albert~R.}\ \bibnamefont
  {Hibbs}}, \ and\ \bibinfo {author} {\bibfnamefont {Daniel~F.}\ \bibnamefont
  {Styer}}} (\bibinfo {year} {2010}),\ \href@noop {} {\emph {\bibinfo {title}
  {Quantum Mechanics and Path Integrals}}}\ (\bibinfo  {publisher} {{Dover
  Publications}},\ \bibinfo {address} {{Mineola, N.Y.}})\BibitemShut {NoStop}%
\bibitem [{\citenamefont {Feynman}\ \emph {et~al.}(1964)\citenamefont
  {Feynman}, \citenamefont {Leighton},\ and\ \citenamefont
  {Sands}}]{1964Feynman}%
  \BibitemOpen
  \bibfield  {author} {\bibinfo {author} {\bibnamefont {Feynman}, \bibfnamefont
  {Richard~P}}, \bibinfo {author} {\bibfnamefont {Robert~B.}\ \bibnamefont
  {Leighton}}, \ and\ \bibinfo {author} {\bibfnamefont {Matthew}\ \bibnamefont
  {Sands}}} (\bibinfo {year} {1964}),\ \href@noop {} {\emph {\bibinfo {title}
  {The {{Feynman}} Lectures on Physics}}}\ (\bibinfo  {publisher}
  {{Addison-Wesley}},\ \bibinfo {address} {{Reading, MA}})\BibitemShut
  {NoStop}%
\bibitem [{\citenamefont {Folland}(1989)}]{1989Folland}%
  \BibitemOpen
  \bibfield  {author} {\bibinfo {author} {\bibnamefont {Folland}, \bibfnamefont
  {G~B}}} (\bibinfo {year} {1989}),\ \href@noop {} {\emph {\bibinfo {title}
  {Harmonic Analysis in Phase Space}}},\ The {{Annals}} of Mathematics Studies
  ; No. 122\ (\bibinfo  {publisher} {{Princeton University Press}},\ \bibinfo
  {address} {{Princeton, N.J.}})\BibitemShut {NoStop}%
\bibitem [{\citenamefont {Gamow}(1964)}]{1964Gamow}%
  \BibitemOpen
  \bibfield  {author} {\bibinfo {author} {\bibnamefont {Gamow}, \bibfnamefont
  {George}}} (\bibinfo {year} {1964}),\ \href@noop {} {\emph {\bibinfo {title}
  {Mr. {{Tompkins}} in {{Wonderland}} : Or {{Stories}} of c, {{G}}, and h}}}\
  (\bibinfo  {publisher} {{Cambridge University Press}},\ \bibinfo {address}
  {{New York}})\BibitemShut {NoStop}%
\bibitem [{\citenamefont {Garay}(1995)}]{1995Garay}%
  \BibitemOpen
  \bibfield  {author} {\bibinfo {author} {\bibnamefont {Garay}, \bibfnamefont
  {L~J}}} (\bibinfo {year} {1995}),\ \bibfield  {title} {\enquote {\bibinfo
  {title} {Quantum-{{Gravity}} and {{Minimum Length}}},}\ }\href@noop {}
  {\bibfield  {journal} {\bibinfo  {journal} {International Journal of Modern
  Physics A}\ }\textbf {\bibinfo {volume} {10}}~(\bibinfo {number} {2}),\
  \bibinfo {pages} {145--165}}\BibitemShut {NoStop}%
\bibitem [{\citenamefont {Ghirardi}\ \emph {et~al.}(1986)\citenamefont
  {Ghirardi}, \citenamefont {Rimini},\ and\ \citenamefont
  {Weber}}]{1986Ghirardi}%
  \BibitemOpen
  \bibfield  {author} {\bibinfo {author} {\bibnamefont {Ghirardi},
  \bibfnamefont {G~C}}, \bibinfo {author} {\bibfnamefont {A.}~\bibnamefont
  {Rimini}}, \ and\ \bibinfo {author} {\bibfnamefont {T.}~\bibnamefont
  {Weber}}} (\bibinfo {year} {1986}),\ \bibfield  {title} {\enquote {\bibinfo
  {title} {Unified dynamics for microscopic and macroscopic systems},}\
  }\href@noop {} {\bibfield  {journal} {\bibinfo  {journal} {Physical Review
  D}\ }\textbf {\bibinfo {volume} {34}}~(\bibinfo {number} {2}),\ \bibinfo
  {pages} {470--491}}\BibitemShut {NoStop}%
\bibitem [{\citenamefont {Gill}\ \emph {et~al.}(2003)\citenamefont {Gill},
  \citenamefont {O'Neill},\ and\ \citenamefont {Besley}}]{2003Gill}%
  \BibitemOpen
  \bibfield  {author} {\bibinfo {author} {\bibnamefont {Gill}, \bibfnamefont
  {Peter~MW}}, \bibinfo {author} {\bibfnamefont {Darragh~P.}\ \bibnamefont
  {O'Neill}}, \ and\ \bibinfo {author} {\bibfnamefont {Nicholas~A.}\
  \bibnamefont {Besley}}} (\bibinfo {year} {2003}),\ \bibfield  {title}
  {\enquote {\bibinfo {title} {Two-electron distribution functions and
  intracules},}\ }\href@noop {} {\bibfield  {journal} {\bibinfo  {journal}
  {Theoretical Chemistry Accounts}\ }\textbf {\bibinfo {volume}
  {109}}~(\bibinfo {number} {5}),\ \bibinfo {pages} {241--250}}\BibitemShut
  {NoStop}%
\bibitem [{\citenamefont {G{\"o}del}(1931)}]{1931Goedel}%
  \BibitemOpen
  \bibfield  {author} {\bibinfo {author} {\bibnamefont {G{\"o}del},
  \bibfnamefont {Kurt}}} (\bibinfo {year} {1931}),\ \bibfield  {title}
  {\enquote {\bibinfo {title} {{{\"U}ber formal unentscheidbare S{\"a}tze der
  Principia Mathematica und verwandter Systeme I}},}\ }\href@noop {} {\bibfield
   {journal} {\bibinfo  {journal} {Monatshefte f{\"u}r Mathematik und Physik}\
  }\textbf {\bibinfo {volume} {38}}~(\bibinfo {number} {1}),\ \bibinfo {pages}
  {173--198}}\BibitemShut {NoStop}%
\bibitem [{\citenamefont {Goldstein}\ \emph {et~al.}(2002)\citenamefont
  {Goldstein}, \citenamefont {Poole},\ and\ \citenamefont
  {Safko}}]{2002Goldstein}%
  \BibitemOpen
  \bibfield  {author} {\bibinfo {author} {\bibnamefont {Goldstein},
  \bibfnamefont {Herbert}}, \bibinfo {author} {\bibfnamefont {Charles.}\
  \bibnamefont {Poole}}, \ and\ \bibinfo {author} {\bibfnamefont {John.}\
  \bibnamefont {Safko}}} (\bibinfo {year} {2002}),\ \href@noop {} {\emph
  {\bibinfo {title} {Classical Mechanics}}}\ (\bibinfo  {publisher} {{Addison
  Wesley}},\ \bibinfo {address} {{San Francisco}})\BibitemShut {NoStop}%
\bibitem [{\citenamefont {Gravel}\ and\ \citenamefont
  {Gauthier}(2011)}]{2011Gravel}%
  \BibitemOpen
  \bibfield  {author} {\bibinfo {author} {\bibnamefont {Gravel}, \bibfnamefont
  {Pierre}}, \ and\ \bibinfo {author} {\bibfnamefont {Claude}\ \bibnamefont
  {Gauthier}}} (\bibinfo {year} {2011}),\ \bibfield  {title} {\enquote
  {\bibinfo {title} {Classical applications of the
  {{Klein}}\textendash{{Gordon}} equation},}\ }\href@noop {} {\bibfield
  {journal} {\bibinfo  {journal} {American Journal of Physics}\ }\textbf
  {\bibinfo {volume} {79}}~(\bibinfo {number} {5}),\ \bibinfo {pages}
  {447--453}}\BibitemShut {NoStop}%
\bibitem [{\citenamefont {Griffiths}(2010)}]{2010Griffiths}%
  \BibitemOpen
  \bibfield  {author} {\bibinfo {author} {\bibnamefont {Griffiths},
  \bibfnamefont {David~J}}} (\bibinfo {year} {2010}),\ \href@noop {} {\emph
  {\bibinfo {title} {Introduction to Elementary Particles}}},\ \bibinfo
  {edition} {2nd}\ ed.,\ Physics Textbook\ (\bibinfo  {publisher}
  {{Wiley-VCH}},\ \bibinfo {address} {{Weinheim}})\BibitemShut {NoStop}%
\bibitem [{\citenamefont {Griffiths}\ and\ \citenamefont
  {Schroeter}(2018)}]{2018Griffiths}%
  \BibitemOpen
  \bibfield  {author} {\bibinfo {author} {\bibnamefont {Griffiths},
  \bibfnamefont {David~J}}, \ and\ \bibinfo {author} {\bibfnamefont
  {Darrell~F.}\ \bibnamefont {Schroeter}}} (\bibinfo {year} {2018}),\
  \href@noop {} {\emph {\bibinfo {title} {Introduction to Quantum
  Mechanics}}},\ \bibinfo {edition} {3rd}\ ed.\ (\bibinfo  {publisher}
  {{Cambridge University Press}})\BibitemShut {NoStop}%
\bibitem [{\citenamefont {Halzen}\ and\ \citenamefont
  {Martin}(1984)}]{1984Halzen}%
  \BibitemOpen
  \bibfield  {author} {\bibinfo {author} {\bibnamefont {Halzen}, \bibfnamefont
  {F}}, \ and\ \bibinfo {author} {\bibfnamefont {Alan~D.}\ \bibnamefont
  {Martin}}} (\bibinfo {year} {1984}),\ \href@noop {} {\emph {\bibinfo {title}
  {Quarks and Leptons : An Introductory Course in Modern Particle Physics}}}\
  (\bibinfo  {publisher} {{Wiley}},\ \bibinfo {address} {{New
  York}})\BibitemShut {NoStop}%
\bibitem [{\citenamefont {Hanbury~Brown}(1974)}]{1974HanburyBrown}%
  \BibitemOpen
  \bibfield  {author} {\bibinfo {author} {\bibnamefont {Hanbury~Brown},
  \bibfnamefont {R}}} (\bibinfo {year} {1974}),\ \href@noop {} {\emph {\bibinfo
  {title} {The Intensity Interferometer; Its Application to Astronomy}}}\
  (\bibinfo  {publisher} {{Taylor and Francis}},\ \bibinfo {address}
  {{London}})\BibitemShut {NoStop}%
\bibitem [{\citenamefont {Hanc}\ \emph {et~al.}(2003)\citenamefont {Hanc},
  \citenamefont {Tuleja},\ and\ \citenamefont {Hancova}}]{2003Hanc}%
  \BibitemOpen
  \bibfield  {author} {\bibinfo {author} {\bibnamefont {Hanc}, \bibfnamefont
  {Jozef}}, \bibinfo {author} {\bibfnamefont {Slavomir}\ \bibnamefont
  {Tuleja}}, \ and\ \bibinfo {author} {\bibfnamefont {Martina}\ \bibnamefont
  {Hancova}}} (\bibinfo {year} {2003}),\ \bibfield  {title} {\enquote {\bibinfo
  {title} {Simple derivation of {{Newtonian}} mechanics from the principle of
  least action},}\ }\href@noop {} {\bibfield  {journal} {\bibinfo  {journal}
  {American Journal of Physics}\ }\textbf {\bibinfo {volume} {71}}~(\bibinfo
  {number} {4}),\ \bibinfo {pages} {386--391}}\BibitemShut {NoStop}%
\bibitem [{\citenamefont {Hanc}\ \emph {et~al.}(2004)\citenamefont {Hanc},
  \citenamefont {Tuleja},\ and\ \citenamefont {Hancova}}]{2004Hanc}%
  \BibitemOpen
  \bibfield  {author} {\bibinfo {author} {\bibnamefont {Hanc}, \bibfnamefont
  {Jozef}}, \bibinfo {author} {\bibfnamefont {Slavomir}\ \bibnamefont
  {Tuleja}}, \ and\ \bibinfo {author} {\bibfnamefont {Martina}\ \bibnamefont
  {Hancova}}} (\bibinfo {year} {2004}),\ \bibfield  {title} {\enquote {\bibinfo
  {title} {Symmetries and conservation laws: {{Consequences}} of {{Noether}}'s
  theorem},}\ }\href@noop {} {\bibfield  {journal} {\bibinfo  {journal}
  {American Journal of Physics}\ }\textbf {\bibinfo {volume} {72}}~(\bibinfo
  {number} {4}),\ \bibinfo {pages} {428--435}}\BibitemShut {NoStop}%
\bibitem [{\citenamefont {Harrison}(2000)}]{2000Harrison}%
  \BibitemOpen
  \bibfield  {author} {\bibinfo {author} {\bibnamefont {Harrison},
  \bibfnamefont {Walter~A}}} (\bibinfo {year} {2000}),\ \href@noop {} {\emph
  {\bibinfo {title} {Applied Quantum Mechanics}}}\ (\bibinfo  {publisher}
  {{World Scientific}},\ \bibinfo {address} {{Singapore ;}})\BibitemShut
  {NoStop}%
\bibitem [{\citenamefont {Hawking}(2002)}]{2002Hawking}%
  \BibitemOpen
  \bibfield  {author} {\bibinfo {author} {\bibnamefont {Hawking}, \bibfnamefont
  {Stephen}}} (\bibinfo {year} {2002}),\ \href@noop {} {\emph {\bibinfo {title}
  {On the Shoulders of Giants : The Great Works of Physics and Astronomy}}}\
  (\bibinfo  {publisher} {{Running Press}},\ \bibinfo {address}
  {{Philadelphia}})\BibitemShut {NoStop}%
\bibitem [{\citenamefont {Hecht}(2005)}]{2005Hecht}%
  \BibitemOpen
  \bibfield  {author} {\bibinfo {author} {\bibnamefont {Hecht}, \bibfnamefont
  {Eugene}}} (\bibinfo {year} {2005}),\ \bibfield  {title} {\enquote {\bibinfo
  {title} {There {{Is No Really Good Definition}} of {{Mass}}},}\ }\href@noop
  {} {\bibfield  {journal} {\bibinfo  {journal} {The Physics Teacher}\ }\textbf
  {\bibinfo {volume} {44}}~(\bibinfo {number} {1}),\ \bibinfo {pages}
  {40--45}}\BibitemShut {NoStop}%
\bibitem [{\citenamefont {Hecht}(2010)}]{2010Hecht}%
  \BibitemOpen
  \bibfield  {author} {\bibinfo {author} {\bibnamefont {Hecht}, \bibfnamefont
  {Eugene}}} (\bibinfo {year} {2010}),\ \bibfield  {title} {\enquote {\bibinfo
  {title} {On {{Defining Mass}}},}\ }\href@noop {} {\bibfield  {journal}
  {\bibinfo  {journal} {The Physics Teacher}\ }\textbf {\bibinfo {volume}
  {49}}~(\bibinfo {number} {1}),\ \bibinfo {pages} {40--44}}\BibitemShut
  {NoStop}%
\bibitem [{\citenamefont {Hecht}(2017)}]{2017Hecht}%
  \BibitemOpen
  \bibfield  {author} {\bibinfo {author} {\bibnamefont {Hecht}, \bibfnamefont
  {Eugene}}} (\bibinfo {year} {2017}),\ \bibfield  {title} {\enquote {\bibinfo
  {title} {A cautionary note on operationally defining force and mass},}\
  }\href@noop {} {\bibfield  {journal} {\bibinfo  {journal} {The Physics
  Teacher}\ }\textbf {\bibinfo {volume} {55}}~(\bibinfo {number} {9}),\
  \bibinfo {pages} {516--517}}\BibitemShut {NoStop}%
\bibitem [{\citenamefont {Hossenfelder}(2013)}]{2013Hossenfelder}%
  \BibitemOpen
  \bibfield  {author} {\bibinfo {author} {\bibnamefont {Hossenfelder},
  \bibfnamefont {Sabine}}} (\bibinfo {year} {2013}),\ \bibfield  {title}
  {\enquote {\bibinfo {title} {Minimal {{Length Scale Scenarios}} for {{Quantum
  Gravity}}},}\ }\href@noop {} {\bibfield  {journal} {\bibinfo  {journal}
  {Living Reviews in Relativity}\ }\textbf {\bibinfo {volume} {16}}~(\bibinfo
  {number} {1}),\ \bibinfo {pages} {2}}\BibitemShut {NoStop}%
\bibitem [{\citenamefont {Inglis}\ \emph {et~al.}(2019)\citenamefont {Inglis},
  \citenamefont {Ullrich},\ and\ \citenamefont {Milton}}]{2019Inglis}%
  \BibitemOpen
  \bibfield  {author} {\bibinfo {author} {\bibnamefont {Inglis}, \bibfnamefont
  {B}}, \bibinfo {author} {\bibfnamefont {J.}~\bibnamefont {Ullrich}}, \ and\
  \bibinfo {author} {\bibfnamefont {M.J.T.}\ \bibnamefont {Milton}}} (\bibinfo
  {year} {2019}),\ \href {www.bipm.org} {\emph {\bibinfo {title} {{{SI
  Brochure}}: {{The International System}} of {{Units}} ({{SI}})}}},\ \bibinfo
  {edition} {9th}\ ed.\ (\bibinfo  {publisher} {{Bureau International des Poids
  et Mesures}})\BibitemShut {NoStop}%
\bibitem [{\citenamefont {Itzykson}\ and\ \citenamefont
  {Zuber}(1980)}]{1980Itzykson}%
  \BibitemOpen
  \bibfield  {author} {\bibinfo {author} {\bibnamefont {Itzykson},
  \bibfnamefont {Claude}}, \ and\ \bibinfo {author} {\bibfnamefont
  {Jean~Bernard}\ \bibnamefont {Zuber}}} (\bibinfo {year} {1980}),\ \href@noop
  {} {\emph {\bibinfo {title} {Quantum Field Theory}}},\ International Series
  in Pure and Applied Physics\ (\bibinfo  {publisher} {{McGraw-Hill
  International Book Co.}},\ \bibinfo {address} {{New York}})\BibitemShut
  {NoStop}%
\bibitem [{\citenamefont {Jackson}(1999)}]{1999Jackson}%
  \BibitemOpen
  \bibfield  {author} {\bibinfo {author} {\bibnamefont {Jackson}, \bibfnamefont
  {John~David}}} (\bibinfo {year} {1999}),\ \href@noop {} {\emph {\bibinfo
  {title} {Classical Electrodynamics}}},\ \bibinfo {edition} {3rd}\ ed.\
  (\bibinfo  {publisher} {{Wiley}},\ \bibinfo {address} {{New
  York}})\BibitemShut {NoStop}%
\bibitem [{\citenamefont {Jacobsen}(2019)}]{2019Jacobsen}%
  \BibitemOpen
  \bibfield  {author} {\bibinfo {author} {\bibnamefont {Jacobsen},
  \bibfnamefont {Chris}}} (\bibinfo {year} {2019}),\ \href@noop {} {\emph
  {\bibinfo {title} {X-Ray {{Microscopy}}}}},\ Advances in {{Microscopy}} and
  {{Microanalysis}}\ (\bibinfo  {publisher} {{Cambridge University Press}},\
  \bibinfo {address} {{Cambridge}})\BibitemShut {NoStop}%
\bibitem [{\citenamefont {Jammer}(1966)}]{1966Jammer}%
  \BibitemOpen
  \bibfield  {author} {\bibinfo {author} {\bibnamefont {Jammer}, \bibfnamefont
  {Max}}} (\bibinfo {year} {1966}),\ \href@noop {} {\emph {\bibinfo {title}
  {The Conceptual Development of Quantum Mechanics}}},\ International Series in
  Pure and Applied Physics\ (\bibinfo  {publisher} {{McGraw-Hill}})\BibitemShut
  {NoStop}%
\bibitem [{\citenamefont {Kia\l{}ka}\ \emph {et~al.}(2019)\citenamefont
  {Kia\l{}ka}, \citenamefont {Stickler}, \citenamefont {Hornberger},
  \citenamefont {Fein}, \citenamefont {Geyer}, \citenamefont {Mairhofer},
  \citenamefont {Gerlich},\ and\ \citenamefont {Arndt}}]{2019Kialka}%
  \BibitemOpen
  \bibfield  {author} {\bibinfo {author} {\bibnamefont {Kia\l{}ka},
  \bibfnamefont {Filip}}, \bibinfo {author} {\bibfnamefont {Benjamin~A.}\
  \bibnamefont {Stickler}}, \bibinfo {author} {\bibfnamefont {Klaus}\
  \bibnamefont {Hornberger}}, \bibinfo {author} {\bibfnamefont {Yaakov~Y.}\
  \bibnamefont {Fein}}, \bibinfo {author} {\bibfnamefont {Philipp}\
  \bibnamefont {Geyer}}, \bibinfo {author} {\bibfnamefont {Lukas}\ \bibnamefont
  {Mairhofer}}, \bibinfo {author} {\bibfnamefont {Stefan}\ \bibnamefont
  {Gerlich}}, \ and\ \bibinfo {author} {\bibfnamefont {Markus}\ \bibnamefont
  {Arndt}}} (\bibinfo {year} {2019}),\ \bibfield  {title} {\enquote {\bibinfo
  {title} {Concepts for long-baseline high-mass matter-wave interferometry},}\
  }\href@noop {} {\bibfield  {journal} {\bibinfo  {journal} {Physica Scripta}\
  }\textbf {\bibinfo {volume} {94}}~(\bibinfo {number} {3}),\ \bibinfo {pages}
  {034001}}\BibitemShut {NoStop}%
\bibitem [{\citenamefont {Kittel}(2005)}]{2005KittelIntroduction}%
  \BibitemOpen
  \bibfield  {author} {\bibinfo {author} {\bibnamefont {Kittel}, \bibfnamefont
  {Charles}}} (\bibinfo {year} {2005}),\ \href@noop {} {\emph {\bibinfo {title}
  {Introduction to Solid State Physics}}},\ \bibinfo {edition} {8th}\ ed.\
  (\bibinfo  {publisher} {{Wiley}},\ \bibinfo {address} {{Hoboken,
  NJ}})\BibitemShut {NoStop}%
\bibitem [{\citenamefont {Kleppner}\ and\ \citenamefont
  {Kolenkow}(2014)}]{2014Kleppner}%
  \BibitemOpen
  \bibfield  {author} {\bibinfo {author} {\bibnamefont {Kleppner},
  \bibfnamefont {Daniel}}, \ and\ \bibinfo {author} {\bibfnamefont {Robert~J.}\
  \bibnamefont {Kolenkow}}} (\bibinfo {year} {2014}),\ \href@noop {} {\emph
  {\bibinfo {title} {An Introduction to Mechanics}}},\ \bibinfo {edition}
  {2nd}\ ed.\ (\bibinfo  {publisher} {{Cambridge University Press}},\ \bibinfo
  {address} {{Cambridge}})\BibitemShut {NoStop}%
\bibitem [{\citenamefont {Lalo{\"e}}(2019)}]{2019Laloe}%
  \BibitemOpen
  \bibfield  {author} {\bibinfo {author} {\bibnamefont {Lalo{\"e}},
  \bibfnamefont {Franck}}} (\bibinfo {year} {2019}),\ \href@noop {} {\emph
  {\bibinfo {title} {Do {{We Really Understand Quantum Mechanics}}?}}},\
  \bibinfo {edition} {2nd}\ ed.\ (\bibinfo  {publisher} {{Cambridge University
  Press}})\BibitemShut {NoStop}%
\bibitem [{\citenamefont {Lamb}\ and\ \citenamefont {Fearn}(1996)}]{1996Lamb}%
  \BibitemOpen
  \bibfield  {author} {\bibinfo {author} {\bibnamefont {Lamb}, \bibfnamefont
  {Willis~E}}, \ and\ \bibinfo {author} {\bibfnamefont {Heidi}\ \bibnamefont
  {Fearn}}} (\bibinfo {year} {1996}),\ \bibfield  {title} {\enquote {\bibinfo
  {title} {Classical theory of measurement: A big step towards the quantum
  theory of measurement},}\ }in\ \href@noop {} {\emph {\bibinfo {booktitle}
  {Amazing Light: A Volume Dedicated To Charles Hard Townes on His 80th
  Birthday}}},\ \bibinfo {editor} {edited by\ \bibinfo {editor} {\bibfnamefont
  {Raymond~Y.}\ \bibnamefont {Chiao}}}\ (\bibinfo  {publisher} {{Springer New
  York}},\ \bibinfo {address} {{New York, NY}})\ pp.\ \bibinfo {pages}
  {373--389}\BibitemShut {NoStop}%
\bibitem [{\citenamefont {Lan}\ \emph {et~al.}(2013)\citenamefont {Lan},
  \citenamefont {Kuan}, \citenamefont {Estey}, \citenamefont {English},
  \citenamefont {Brown}, \citenamefont {Hohensee},\ and\ \citenamefont
  {M{\"u}ller}}]{2013Lan}%
  \BibitemOpen
  \bibfield  {author} {\bibinfo {author} {\bibnamefont {Lan}, \bibfnamefont
  {Shau-Yu}}, \bibinfo {author} {\bibfnamefont {Pei-Chen}\ \bibnamefont
  {Kuan}}, \bibinfo {author} {\bibfnamefont {Brian}\ \bibnamefont {Estey}},
  \bibinfo {author} {\bibfnamefont {Damon}\ \bibnamefont {English}}, \bibinfo
  {author} {\bibfnamefont {Justin~M.}\ \bibnamefont {Brown}}, \bibinfo {author}
  {\bibfnamefont {Michael~A.}\ \bibnamefont {Hohensee}}, \ and\ \bibinfo
  {author} {\bibfnamefont {Holger}\ \bibnamefont {M{\"u}ller}}} (\bibinfo
  {year} {2013}),\ \bibfield  {title} {\enquote {\bibinfo {title} {A {{Clock
  Directly Linking Time}} to a {{Particle}}'s {{Mass}}},}\ }\href@noop {}
  {\bibfield  {journal} {\bibinfo  {journal} {Science}\ }\textbf {\bibinfo
  {volume} {339}}~(\bibinfo {number} {6119}),\ \bibinfo {pages}
  {554--557}}\BibitemShut {NoStop}%
\bibitem [{\citenamefont {Landau}\ and\ \citenamefont
  {Lifshitz}(1981)}]{1981Landau}%
  \BibitemOpen
  \bibfield  {author} {\bibinfo {author} {\bibnamefont {Landau}, \bibfnamefont
  {L~D}}, \ and\ \bibinfo {author} {\bibfnamefont {E.~M.}\ \bibnamefont
  {Lifshitz}}} (\bibinfo {year} {1981}),\ \href@noop {} {\emph {\bibinfo
  {title} {Quantum Mechanics : Non-Relativistic Theory}}},\ \bibinfo {edition}
  {3rd}\ ed.,\ Course of Theoretical Physics ; {Vol.} 3\ (\bibinfo  {publisher}
  {{Pergamon}},\ \bibinfo {address} {{Oxford}})\BibitemShut {NoStop}%
\bibitem [{\citenamefont {Landau}\ \emph {et~al.}(1976)\citenamefont {Landau},
  \citenamefont {Lifshitz}, \citenamefont {Sykes},\ and\ \citenamefont
  {Bell}}]{1976Landau}%
  \BibitemOpen
  \bibfield  {author} {\bibinfo {author} {\bibnamefont {Landau}, \bibfnamefont
  {Lev~Davidovitch}}, \bibinfo {author} {\bibfnamefont {Evgenii~Mikhailovich}\
  \bibnamefont {Lifshitz}}, \bibinfo {author} {\bibfnamefont {John~Bradbury}\
  \bibnamefont {Sykes}}, \ and\ \bibinfo {author} {\bibfnamefont
  {John~Stewart}\ \bibnamefont {Bell}}} (\bibinfo {year} {1976}),\ \href@noop
  {} {\emph {\bibinfo {title} {Mechanics}}},\ \bibinfo {edition} {3rd}\ ed.,\
  Course of {{Theorical Physics}} ; {V}ol 1\ (\bibinfo  {publisher}
  {{Butterworth-Heinemann}},\ \bibinfo {address} {{Oxford}})\BibitemShut
  {NoStop}%
\bibitem [{\citenamefont {Mermin}(1985)}]{1985Mermin}%
  \BibitemOpen
  \bibfield  {author} {\bibinfo {author} {\bibnamefont {Mermin}, \bibfnamefont
  {N~David}}} (\bibinfo {year} {1985}),\ \bibfield  {title} {\enquote {\bibinfo
  {title} {Is the {{Moon There When Nobody Looks}}? {{Reality}} and the
  {{Quantum Theory}}},}\ }\href@noop {} {\bibfield  {journal} {\bibinfo
  {journal} {Physics Today}\ }\textbf {\bibinfo {volume} {38}}~(\bibinfo
  {number} {4}),\ \bibinfo {pages} {38}}\BibitemShut {NoStop}%
\bibitem [{\citenamefont {Messiah}(2014)}]{2014Messiah}%
  \BibitemOpen
  \bibfield  {author} {\bibinfo {author} {\bibnamefont {Messiah}, \bibfnamefont
  {Albert}}} (\bibinfo {year} {2014}),\ \href@noop {} {\emph {\bibinfo {title}
  {Quantum Mechanics : Two Volumes Bound as One}}},\ Dover Books on Physics\
  (\bibinfo  {publisher} {{Dover}},\ \bibinfo {address} {{Mineola,
  NY}})\BibitemShut {NoStop}%
\bibitem [{\citenamefont {Misner}\ \emph {et~al.}(1973)\citenamefont {Misner},
  \citenamefont {Thorne},\ and\ \citenamefont {Wheeler}}]{1973Misner}%
  \BibitemOpen
  \bibfield  {author} {\bibinfo {author} {\bibnamefont {Misner}, \bibfnamefont
  {Charles~W}}, \bibinfo {author} {\bibfnamefont {Kip~S.}\ \bibnamefont
  {Thorne}}, \ and\ \bibinfo {author} {\bibfnamefont {John~Archibald}\
  \bibnamefont {Wheeler}}} (\bibinfo {year} {1973}),\ \href@noop {} {\emph
  {\bibinfo {title} {Gravitation}}}\ (\bibinfo  {publisher} {{W.H. Freeman and
  Company}},\ \bibinfo {address} {{New York}})\BibitemShut {NoStop}%
\bibitem [{\citenamefont {von Neumann}(2018)}]{2018Neumann}%
  \BibitemOpen
  \bibfield  {author} {\bibinfo {author} {\bibnamefont {von Neumann},
  \bibfnamefont {John}}} (\bibinfo {year} {2018}),\ \href@noop {} {\emph
  {\bibinfo {title} {Mathematical Foundations of Quantum Mechanics}}},\
  \bibinfo {edition} {{N}ew}\ ed.,\ edited by\ \bibinfo {editor} {\bibfnamefont
  {Nicholas~A.}\ \bibnamefont {Wheeler}}\ (\bibinfo  {publisher} {{Princeton
  University Press}},\ \bibinfo {address} {{Princeton}})\BibitemShut {NoStop}%
\bibitem [{\citenamefont {Newton}\ \emph {et~al.}(2016)\citenamefont {Newton},
  \citenamefont {Budenz}, \citenamefont {Cohen},\ and\ \citenamefont
  {Whitman}}]{2016Newton}%
  \BibitemOpen
  \bibfield  {author} {\bibinfo {author} {\bibnamefont {Newton}, \bibfnamefont
  {Isaac}}, \bibinfo {author} {\bibfnamefont {Julia}\ \bibnamefont {Budenz}},
  \bibinfo {author} {\bibfnamefont {I.~Bernard.}\ \bibnamefont {Cohen}}, \ and\
  \bibinfo {author} {\bibfnamefont {Anne~Miller}\ \bibnamefont {Whitman}}}
  (\bibinfo {year} {2016}),\ \href@noop {} {\emph {\bibinfo {title} {The
  Principia : Mathematical Principles of Natural Philosophy}}}\ (\bibinfo
  {publisher} {{University of California Press}},\ \bibinfo {address}
  {{Berkerley, California}})\BibitemShut {NoStop}%
\bibitem [{\citenamefont {Newton}\ and\ \citenamefont
  {Wigner}(1949)}]{1949Newton}%
  \BibitemOpen
  \bibfield  {author} {\bibinfo {author} {\bibnamefont {Newton}, \bibfnamefont
  {T~D}}, \ and\ \bibinfo {author} {\bibfnamefont {E.~P.}\ \bibnamefont
  {Wigner}}} (\bibinfo {year} {1949}),\ \bibfield  {title} {\enquote {\bibinfo
  {title} {Localized {{States}} for {{Elementary Systems}}},}\ }\href@noop {}
  {\bibfield  {journal} {\bibinfo  {journal} {Reviews of Modern Physics}\
  }\textbf {\bibinfo {volume} {21}}~(\bibinfo {number} {3}),\ \bibinfo {pages}
  {400--406}}\BibitemShut {NoStop}%
\bibitem [{\citenamefont {Ogborn}\ \emph {et~al.}(2006)\citenamefont {Ogborn},
  \citenamefont {Hanc},\ and\ \citenamefont {Taylor}}]{2006Ogborn}%
  \BibitemOpen
  \bibfield  {author} {\bibinfo {author} {\bibnamefont {Ogborn}, \bibfnamefont
  {Jon}}, \bibinfo {author} {\bibfnamefont {Jozef}\ \bibnamefont {Hanc}}, \
  and\ \bibinfo {author} {\bibfnamefont {Edwin~F.}\ \bibnamefont {Taylor}}}
  (\bibinfo {year} {2006}),\ \bibfield  {title} {\enquote {\bibinfo {title}
  {Action on {{Stage}}: {{Historical Introduction}}},}\ }in\ \href@noop {}
  {\emph {\bibinfo {booktitle} {The {{Girep Conference} 2006}}}}\ (\bibinfo
  {publisher} {{Universiteit van Amsterdam}})\BibitemShut {NoStop}%
\bibitem [{\citenamefont {Ogborn}\ and\ \citenamefont
  {Taylor}(2005)}]{2005Ogborn}%
  \BibitemOpen
  \bibfield  {author} {\bibinfo {author} {\bibnamefont {Ogborn}, \bibfnamefont
  {Jon}}, \ and\ \bibinfo {author} {\bibfnamefont {Edwin~F.}\ \bibnamefont
  {Taylor}}} (\bibinfo {year} {2005}),\ \bibfield  {title} {\enquote {\bibinfo
  {title} {Quantum physics explains {{Newton}}'s laws of motion},}\ }\href@noop
  {} {\bibfield  {journal} {\bibinfo  {journal} {Physics Education}\ }\textbf
  {\bibinfo {volume} {40}}~(\bibinfo {number} {1}),\ \bibinfo {pages}
  {26--34}}\BibitemShut {NoStop}%
\bibitem [{\citenamefont {Omn{\`e}s}(1971)}]{1971Omnes}%
  \BibitemOpen
  \bibfield  {author} {\bibinfo {author} {\bibnamefont {Omn{\`e}s},
  \bibfnamefont {Roland}}} (\bibinfo {year} {1971}),\ \href@noop {} {\emph
  {\bibinfo {title} {Introduction to Particle Physics.}}}\ (\bibinfo
  {publisher} {{Wiley-Interscience}},\ \bibinfo {address}
  {{London}})\BibitemShut {NoStop}%
\bibitem [{\citenamefont {Pais}(2005)}]{2005Pais}%
  \BibitemOpen
  \bibfield  {author} {\bibinfo {author} {\bibnamefont {Pais}, \bibfnamefont
  {Abraham}}} (\bibinfo {year} {2005}),\ \href@noop {} {\emph {\bibinfo {title}
  {"{{Subtle}} Is the {{Lord}}-- " : The Science and the Life of {{Albert
  Einstein}}}}}\ (\bibinfo  {publisher} {{Oxford University Press}},\ \bibinfo
  {address} {{Oxford; New York}})\BibitemShut {NoStop}%
\bibitem [{\citenamefont {Pease}(2013)}]{2013Pease}%
  \BibitemOpen
  \bibfield  {author} {\bibinfo {author} {\bibnamefont {Pease}, \bibfnamefont
  {Roland}}} (\bibinfo {year} {2013}),\ \bibfield  {title} {\enquote {\bibinfo
  {title} {The time? {{About}} a quarter past a kilogram},}\ }\href@noop {}
  {\bibinfo  {journal} {Nature News}\ }\BibitemShut {NoStop}%
\bibitem [{\citenamefont {Peskin}\ and\ \citenamefont
  {Schroeder}(1995)}]{1995Peskin}%
  \BibitemOpen
\bibfield  {journal} {  }\bibfield  {author} {\bibinfo {author} {\bibnamefont
  {Peskin}, \bibfnamefont {Michael~Edward}}, \ and\ \bibinfo {author}
  {\bibfnamefont {Daniel~V.}\ \bibnamefont {Schroeder}}} (\bibinfo {year}
  {1995}),\ \href@noop {} {\emph {\bibinfo {title} {An Introduction to Quantum
  Field Theory}}},\ Advanced Book Program\ (\bibinfo  {publisher}
  {{Addison-Wesley}},\ \bibinfo {address} {{Reading, Mass.}})\BibitemShut
  {NoStop}%
\bibitem [{\citenamefont {Proud}\ and\ \citenamefont
  {Pearson}(2010)}]{2010Proud}%
  \BibitemOpen
  \bibfield  {author} {\bibinfo {author} {\bibnamefont {Proud}, \bibfnamefont
  {Adam~J}}, \ and\ \bibinfo {author} {\bibfnamefont {Jason~K.}\ \bibnamefont
  {Pearson}}} (\bibinfo {year} {2010}),\ \bibfield  {title} {\enquote {\bibinfo
  {title} {A simultaneous probability density for the intracule and extracule
  coordinates},}\ }\href@noop {} {\bibfield  {journal} {\bibinfo  {journal}
  {The Journal of Chemical Physics}\ }\textbf {\bibinfo {volume}
  {133}}~(\bibinfo {number} {13}),\ \bibinfo {pages} {134113}}\BibitemShut
  {NoStop}%
\bibitem [{\citenamefont {Pryce}\ and\ \citenamefont
  {Chapman}(1948)}]{1948Pryce}%
  \BibitemOpen
  \bibfield  {author} {\bibinfo {author} {\bibnamefont {Pryce}, \bibfnamefont
  {Maurice Henry~Lecorney}}, \ and\ \bibinfo {author} {\bibfnamefont {Sydney}\
  \bibnamefont {Chapman}}} (\bibinfo {year} {1948}),\ \bibfield  {title}
  {\enquote {\bibinfo {title} {The mass-centre in the restricted theory of
  relativity and its connexion with the quantum theory of elementary
  particles},}\ }\href@noop {} {\bibfield  {journal} {\bibinfo  {journal}
  {Proceedings of the Royal Society of London. Series A. Mathematical and
  Physical Sciences}\ }\textbf {\bibinfo {volume} {195}}~(\bibinfo {number}
  {1040}),\ \bibinfo {pages} {62--81}}\BibitemShut {NoStop}%
\bibitem [{\citenamefont {Purcell}(1956)}]{1956Purcell}%
  \BibitemOpen
  \bibfield  {author} {\bibinfo {author} {\bibnamefont {Purcell}, \bibfnamefont
  {E~M}}} (\bibinfo {year} {1956}),\ \bibfield  {title} {\enquote {\bibinfo
  {title} {The {{Question}} of {{Correlation}} between {{Photons}} in
  {{Coherent Light Rays}}},}\ }\href@noop {} {\bibfield  {journal} {\bibinfo
  {journal} {Nature}\ }\textbf {\bibinfo {volume} {178}}~(\bibinfo {number}
  {4548}),\ \bibinfo {pages} {1449--1450}}\BibitemShut {NoStop}%
\bibitem [{\citenamefont {Ralston}(2012)}]{2012Ralston}%
  \BibitemOpen
  \bibfield  {author} {\bibinfo {author} {\bibnamefont {Ralston}, \bibfnamefont
  {John~P}}} (\bibinfo {year} {2012}),\ \bibfield  {title} {\enquote {\bibinfo
  {title} {Quantum {{Theory}} without {{Planck}}'s {{Constant}}},}\ }\href@noop
  {} {\bibinfo  {journal} {arXiv:1203.5557 [hep-ph]}\ }\BibitemShut {NoStop}%
\bibitem [{\citenamefont {Ralston}(2013)}]{2013Ralston}%
  \BibitemOpen
\bibfield  {journal} {  }\bibfield  {author} {\bibinfo {author} {\bibnamefont
  {Ralston}, \bibfnamefont {John~P}}} (\bibinfo {year} {2013}),\ \bibfield
  {title} {\enquote {\bibinfo {title} {Revising your world-view of the
  fundamental constants},}\ }in\ \href@noop {} {\emph {\bibinfo {booktitle}
  {The {{Nature}} of {{Light}}: {{What}} Are {{Photons}}? {{V}}}}},\ Vol.\
  \bibinfo {volume} {8832}\ (\bibinfo  {publisher} {{International Society for
  Optics and Photonics}})\ p.\ \bibinfo {pages} {883216}\BibitemShut {NoStop}%
\bibitem [{\citenamefont {Reimer}\ and\ \citenamefont
  {Kohl}(2008)}]{2008Reimer}%
  \BibitemOpen
  \bibfield  {author} {\bibinfo {author} {\bibnamefont {Reimer}, \bibfnamefont
  {Ludwig}}, \ and\ \bibinfo {author} {\bibfnamefont {Helmut.}\ \bibnamefont
  {Kohl}}} (\bibinfo {year} {2008}),\ \href@noop {} {\emph {\bibinfo {title}
  {Transmission Electron Microscopy : Physics of Image Formation}}}\ (\bibinfo
  {publisher} {{Springer}},\ \bibinfo {address} {{Berlin}})\BibitemShut
  {NoStop}%
\bibitem [{\citenamefont {Sakurai}(1967)}]{1967Sakurai}%
  \BibitemOpen
  \bibfield  {author} {\bibinfo {author} {\bibnamefont {Sakurai}, \bibfnamefont
  {J~J}}} (\bibinfo {year} {1967}),\ \href@noop {} {\emph {\bibinfo {title}
  {Advanced Quantum Mechanics}}},\ Addison-{{Wesley}} Series in Advanced
  Physics\ (\bibinfo  {publisher} {{Addison-Wesley Pub. Co.}},\ \bibinfo
  {address} {{Reading, Mass.}})\BibitemShut {NoStop}%
\bibitem [{\citenamefont {Sakurai}\ and\ \citenamefont
  {Napolitano}(2017)}]{2017Sakurai}%
  \BibitemOpen
  \bibfield  {author} {\bibinfo {author} {\bibnamefont {Sakurai}, \bibfnamefont
  {J~J}}, \ and\ \bibinfo {author} {\bibfnamefont {Jim}\ \bibnamefont
  {Napolitano}}} (\bibinfo {year} {2017}),\ \href@noop {} {\emph {\bibinfo
  {title} {Modern Quantum Mechanics}}},\ \bibinfo {edition} {2nd}\ ed.\
  (\bibinfo  {publisher} {{Cambridge University Press}},\ \bibinfo {address}
  {{Cambridge, United Kingdom}})\BibitemShut {NoStop}%
\bibitem [{\citenamefont {Savitt}(1997)}]{1997Savitt}%
  \BibitemOpen
  \bibinfo {editor} {\bibnamefont {Savitt}, \bibfnamefont {Steven~Frederick}},\
  Ed. (\bibinfo {year} {1997}),\ \href@noop {} {\emph {\bibinfo {title} {Time's
  Arrows Today : Recent Physical and Philosophical Work on the Direction of
  Time}}}\ (\bibinfo  {publisher} {{Cambridge University Press}},\ \bibinfo
  {address} {{Cambridge ;}})\BibitemShut {NoStop}%
\bibitem [{\citenamefont {Saxon}(1968)}]{1968Saxon}%
  \BibitemOpen
  \bibfield  {author} {\bibinfo {author} {\bibnamefont {Saxon}, \bibfnamefont
  {David~S}}} (\bibinfo {year} {1968}),\ \href@noop {} {\emph {\bibinfo {title}
  {Elementary Quantum Mechanics}}},\ Holden-{{Day}} Series in Physics\
  (\bibinfo  {publisher} {{Holden-Day}},\ \bibinfo {address} {{San
  Francisco}})\BibitemShut {NoStop}%
\bibitem [{\citenamefont {Schiff}(1968)}]{1968Schiff}%
  \BibitemOpen
  \bibfield  {author} {\bibinfo {author} {\bibnamefont {Schiff}, \bibfnamefont
  {Leonard~I}}} (\bibinfo {year} {1968}),\ \href@noop {} {\emph {\bibinfo
  {title} {Quantum Mechanics}}},\ \bibinfo {edition} {3rd}\ ed.,\ International
  Series in Pure and Applied Physics\ (\bibinfo  {publisher} {{McGraw-Hill}},\
  \bibinfo {address} {{New York}})\BibitemShut {NoStop}%
\bibitem [{\citenamefont {Schleich}\ \emph {et~al.}(2013)\citenamefont
  {Schleich}, \citenamefont {Greenberger},\ and\ \citenamefont
  {Rasel}}]{2013Schleich}%
  \BibitemOpen
  \bibfield  {author} {\bibinfo {author} {\bibnamefont {Schleich},
  \bibfnamefont {Wolfgang~P}}, \bibinfo {author} {\bibfnamefont {Daniel~M.}\
  \bibnamefont {Greenberger}}, \ and\ \bibinfo {author} {\bibfnamefont
  {Ernst~M.}\ \bibnamefont {Rasel}}} (\bibinfo {year} {2013}),\ \bibfield
  {title} {\enquote {\bibinfo {title} {A representation-free description of the
  {{Kasevich}}\textendash{{Chu}} interferometer: A resolution of the redshift
  controversy},}\ }\href@noop {} {\bibfield  {journal} {\bibinfo  {journal}
  {New Journal of Physics}\ }\textbf {\bibinfo {volume} {15}}~(\bibinfo
  {number} {1}),\ \bibinfo {pages} {013007}}\BibitemShut {NoStop}%
\bibitem [{\citenamefont {Schulman}(1997)}]{1997Schulman}%
  \BibitemOpen
  \bibfield  {author} {\bibinfo {author} {\bibnamefont {Schulman},
  \bibfnamefont {Lawrence~S}}} (\bibinfo {year} {1997}),\ \href@noop {} {\emph
  {\bibinfo {title} {Time's Arrows and Quantum Measurement}}}\ (\bibinfo
  {publisher} {{Cambridge University Press}},\ \bibinfo {address} {{New
  York}})\BibitemShut {NoStop}%
\bibitem [{\citenamefont {Schwarzenbach}(1996)}]{1996Schwarzenbach}%
  \BibitemOpen
  \bibfield  {author} {\bibinfo {author} {\bibnamefont {Schwarzenbach},
  \bibfnamefont {Dieter}}} (\bibinfo {year} {1996}),\ \href@noop {} {\emph
  {\bibinfo {title} {Crystallography}}}\ (\bibinfo  {publisher} {{John
  Wiley}},\ \bibinfo {address} {{New York}})\BibitemShut {NoStop}%
\bibitem [{\citenamefont {Shapere}\ and\ \citenamefont
  {Wilczek}(1989)}]{1989Shapere}%
  \BibitemOpen
  \bibfield  {author} {\bibinfo {author} {\bibnamefont {Shapere}, \bibfnamefont
  {Alfred}}, \ and\ \bibinfo {author} {\bibfnamefont {Frank.}\ \bibnamefont
  {Wilczek}}} (\bibinfo {year} {1989}),\ \href@noop {} {\emph {\bibinfo {title}
  {Geometric Phases in Physics}}},\ Advanced Series in Mathematical Physics ;
  v. 5\ (\bibinfo  {publisher} {{World Scientific}},\ \bibinfo {address}
  {{Singapore}})\BibitemShut {NoStop}%
\bibitem [{\citenamefont {Stein}\ and\ \citenamefont
  {Shakarchi}(2003)}]{2003Stein}%
  \BibitemOpen
  \bibfield  {author} {\bibinfo {author} {\bibnamefont {Stein}, \bibfnamefont
  {Elias~M}}, \ and\ \bibinfo {author} {\bibfnamefont {Rami.}\ \bibnamefont
  {Shakarchi}}} (\bibinfo {year} {2003}),\ \href@noop {} {\emph {\bibinfo
  {title} {Fourier Analysis : An Introduction}}},\ Princeton Lectures in
  Analysis ; 1\ (\bibinfo  {publisher} {{Princeton University Press}},\
  \bibinfo {address} {{Princeton}})\BibitemShut {NoStop}%
\bibitem [{\citenamefont {Stenger}\ and\ \citenamefont
  {Ullrich}(2016)}]{2016Stenger}%
  \BibitemOpen
  \bibfield  {author} {\bibinfo {author} {\bibnamefont {Stenger}, \bibfnamefont
  {J}}, \ and\ \bibinfo {author} {\bibfnamefont {J.H.}\ \bibnamefont
  {Ullrich}}} (\bibinfo {year} {2016}),\ \bibfield  {title} {\enquote {\bibinfo
  {title} {Units {{Based}} on {{Constants}}: {{The Redefinition}} of the
  {{International System}} of {{Units}}},}\ }\href@noop {} {\bibfield
  {journal} {\bibinfo  {journal} {Annual Review of Condensed Matter Physics}\
  }\textbf {\bibinfo {volume} {7}}~(\bibinfo {number} {1}),\ \bibinfo {pages}
  {35--59}}\BibitemShut {NoStop}%
\bibitem [{\citenamefont {Strunk}\ and\ \citenamefont
  {White}(2009)}]{2009Strunk}%
  \BibitemOpen
  \bibfield  {author} {\bibinfo {author} {\bibnamefont {Strunk}, \bibfnamefont
  {William}}, \ and\ \bibinfo {author} {\bibfnamefont {E.~B.}\ \bibnamefont
  {White}}} (\bibinfo {year} {2009}),\ \href@noop {} {\emph {\bibinfo {title}
  {The Elements of Style}}},\ \bibinfo {edition} {{F}iftieth anniversary}\ ed.\
  (\bibinfo  {publisher} {{Pearson Longman}},\ \bibinfo {address} {{New
  York}})\BibitemShut {NoStop}%
\bibitem [{\citenamefont {Styer}\ \emph {et~al.}(2002)\citenamefont {Styer},
  \citenamefont {Balkin}, \citenamefont {Becker}, \citenamefont {Burns},
  \citenamefont {Dudley}, \citenamefont {Forth}, \citenamefont {Gaumer},
  \citenamefont {Kramer}, \citenamefont {Oertel}, \citenamefont {Park},
  \citenamefont {Rinkoski}, \citenamefont {Smith},\ and\ \citenamefont
  {Wotherspoon}}]{2002Styer}%
  \BibitemOpen
  \bibfield  {author} {\bibinfo {author} {\bibnamefont {Styer}, \bibfnamefont
  {Daniel~F}}, \bibinfo {author} {\bibfnamefont {Miranda~S.}\ \bibnamefont
  {Balkin}}, \bibinfo {author} {\bibfnamefont {Kathryn~M.}\ \bibnamefont
  {Becker}}, \bibinfo {author} {\bibfnamefont {Matthew~R.}\ \bibnamefont
  {Burns}}, \bibinfo {author} {\bibfnamefont {Christopher~E.}\ \bibnamefont
  {Dudley}}, \bibinfo {author} {\bibfnamefont {Scott~T.}\ \bibnamefont
  {Forth}}, \bibinfo {author} {\bibfnamefont {Jeremy~S.}\ \bibnamefont
  {Gaumer}}, \bibinfo {author} {\bibfnamefont {Mark~A.}\ \bibnamefont
  {Kramer}}, \bibinfo {author} {\bibfnamefont {David~C.}\ \bibnamefont
  {Oertel}}, \bibinfo {author} {\bibfnamefont {Leonard~H.}\ \bibnamefont
  {Park}}, \bibinfo {author} {\bibfnamefont {Marie~T.}\ \bibnamefont
  {Rinkoski}}, \bibinfo {author} {\bibfnamefont {Clait~T.}\ \bibnamefont
  {Smith}}, \ and\ \bibinfo {author} {\bibfnamefont {Timothy~D.}\ \bibnamefont
  {Wotherspoon}}} (\bibinfo {year} {2002}),\ \bibfield  {title} {\enquote
  {\bibinfo {title} {Nine formulations of quantum mechanics},}\ }\href@noop {}
  {\bibfield  {journal} {\bibinfo  {journal} {American Journal of Physics}\
  }\textbf {\bibinfo {volume} {70}}~(\bibinfo {number} {3}),\ \bibinfo {pages}
  {288--297}}\BibitemShut {NoStop}%
\bibitem [{\citenamefont {Symon}(1971)}]{1971Symon}%
  \BibitemOpen
  \bibfield  {author} {\bibinfo {author} {\bibnamefont {Symon}, \bibfnamefont
  {Keith~R}}} (\bibinfo {year} {1971}),\ \href@noop {} {\emph {\bibinfo {title}
  {Mechanics}}},\ \bibinfo {edition} {3rd}\ ed.,\ Addison-{{Wesley}} Series in
  Physics\ (\bibinfo  {publisher} {{Addison-Wesley Pub. Co.}},\ \bibinfo
  {address} {{Reading, Mass.}})\BibitemShut {NoStop}%
\bibitem [{\citenamefont {{'t Hooft}}(2018)}]{2018tHooft}%
  \BibitemOpen
  \bibfield  {author} {\bibinfo {author} {\bibnamefont {{'t Hooft}},
  \bibfnamefont {Gerard}}} (\bibinfo {year} {2018}),\ \bibfield  {title}
  {\enquote {\bibinfo {title} {Time, the {{Arrow}} of {{Time}}, and {{Quantum
  Mechanics}}},}\ }\href@noop {} {\bibfield  {journal} {\bibinfo  {journal}
  {Frontiers in Physics}\ }\textbf {\bibinfo {volume} {6}}}\BibitemShut
  {NoStop}%
\bibitem [{\citenamefont {Tanabashi}(2018)}]{2018ParticleDataGroup}%
  \BibitemOpen
  \bibfield  {author} {\bibinfo {author} {\bibnamefont {Tanabashi},
  \bibfnamefont {M~\emph{et al} ({P}article {D}ata~{G}roup)}}} (\bibinfo {year}
  {2018}),\ \bibfield  {title} {\enquote {\bibinfo {title} {Review of
  {{Particle Physics}}},}\ }\href@noop {} {\bibfield  {journal} {\bibinfo
  {journal} {Physical Review D}\ }\textbf {\bibinfo {volume} {98}}~(\bibinfo
  {number} {3}),\ \bibinfo {pages} {030001}}\BibitemShut {NoStop}%
\bibitem [{\citenamefont {Tarski}(1936)}]{1936Tarski}%
  \BibitemOpen
  \bibfield  {author} {\bibinfo {author} {\bibnamefont {Tarski}, \bibfnamefont
  {Alfred}}} (\bibinfo {year} {1936}),\ \bibfield  {title} {\enquote {\bibinfo
  {title} {Der {{Wahrheitsbegriff}} in den formalisierten {{Sprachen}}},}\
  }\href@noop {} {\bibfield  {journal} {\bibinfo  {journal} {Studia
  Philosophica}\ }\textbf {\bibinfo {volume} {1}},\ \bibinfo {pages}
  {261--405}}\BibitemShut {NoStop}%
\bibitem [{\citenamefont {Taylor}\ and\ \citenamefont
  {Wheeler}(1992)}]{1992Taylor}%
  \BibitemOpen
  \bibfield  {author} {\bibinfo {author} {\bibnamefont {Taylor}, \bibfnamefont
  {Edwin~F}}, \ and\ \bibinfo {author} {\bibfnamefont {John~Archibald}\
  \bibnamefont {Wheeler}}} (\bibinfo {year} {1992}),\ \href@noop {} {\emph
  {\bibinfo {title} {Spacetime Physics : Introduction to Special Relativity}}}\
  (\bibinfo  {publisher} {{W.H. Freeman}},\ \bibinfo {address} {{New
  York}})\BibitemShut {NoStop}%
\bibitem [{\citenamefont {Thornton}\ and\ \citenamefont
  {Marion}(2004)}]{2004Thornton}%
  \BibitemOpen
  \bibfield  {author} {\bibinfo {author} {\bibnamefont {Thornton},
  \bibfnamefont {Stephen~T}}, \ and\ \bibinfo {author} {\bibfnamefont
  {Jerry~B.}\ \bibnamefont {Marion}}} (\bibinfo {year} {2004}),\ \href@noop {}
  {\emph {\bibinfo {title} {Classical Dynamics of Particles and Systems.}}}\
  (\bibinfo  {publisher} {{Brooks/Cole}},\ \bibinfo {address} {{Belmont,
  CA}})\BibitemShut {NoStop}%
\bibitem [{\citenamefont {Wichmann}(1971)}]{1971Wichmann}%
  \BibitemOpen
  \bibfield  {author} {\bibinfo {author} {\bibnamefont {Wichmann},
  \bibfnamefont {Eyvind~H}}} (\bibinfo {year} {1971}),\ \href@noop {} {\emph
  {\bibinfo {title} {Quantum Physics}}},\ \bibinfo {series} {Berkeley {{Physics
  Course}}}, Vol.~\bibinfo {volume} {4}\ (\bibinfo  {publisher} {{McGraw-Hill
  Book Company}},\ \bibinfo {address} {{New York}})\BibitemShut {NoStop}%
\bibitem [{\citenamefont {Wolf}\ \emph {et~al.}(2012)\citenamefont {Wolf},
  \citenamefont {Blanchet}, \citenamefont {Bord{\'e}}, \citenamefont {Reynaud},
  \citenamefont {Salomon},\ and\ \citenamefont {{Cohen-Tannoudji}}}]{2012Wolf}%
  \BibitemOpen
  \bibfield  {author} {\bibinfo {author} {\bibnamefont {Wolf}, \bibfnamefont
  {Peter}}, \bibinfo {author} {\bibfnamefont {Luc}\ \bibnamefont {Blanchet}},
  \bibinfo {author} {\bibfnamefont {Christian~J.}\ \bibnamefont {Bord{\'e}}},
  \bibinfo {author} {\bibfnamefont {Serge}\ \bibnamefont {Reynaud}}, \bibinfo
  {author} {\bibfnamefont {Christophe}\ \bibnamefont {Salomon}}, \ and\
  \bibinfo {author} {\bibfnamefont {Claude}\ \bibnamefont {{Cohen-Tannoudji}}}}
  (\bibinfo {year} {2012}),\ \bibfield  {title} {\enquote {\bibinfo {title}
  {Reply to comment on: `{{Does}} an atom interferometer test the gravitational
  redshift at the {{Compton}} frequency?'},}\ }\href@noop {} {\bibfield
  {journal} {\bibinfo  {journal} {Classical and Quantum Gravity}\ }\textbf
  {\bibinfo {volume} {29}}~(\bibinfo {number} {4}),\ \bibinfo {pages}
  {048002}}\BibitemShut {NoStop}%
\bibitem [{\citenamefont {Young}\ \emph {et~al.}(2020)\citenamefont {Young},
  \citenamefont {Freedman},\ and\ \citenamefont {Ford}}]{2020Young}%
  \BibitemOpen
  \bibfield  {author} {\bibinfo {author} {\bibnamefont {Young}, \bibfnamefont
  {Hugh~D}}, \bibinfo {author} {\bibfnamefont {Roger~A.}\ \bibnamefont
  {Freedman}}, \ and\ \bibinfo {author} {\bibfnamefont {Albert~Lewis.}\
  \bibnamefont {Ford}}} (\bibinfo {year} {2020}),\ \href@noop {} {\emph
  {\bibinfo {title} {Sears and {{Zemansky}}'s University Physics with Modern
  Physics}}},\ \bibinfo {edition} {15th}\ ed.\ (\bibinfo  {publisher}
  {{Pearson}},\ \bibinfo {address} {{Hoboken (N. J.)}})\BibitemShut {NoStop}%
\end{thebibliography}%
%

\end{document}